\newcommand{\changes}[1]{\textcolor{black}{#1}}

\documentclass[sigconf]{acmart}
\usepackage{enumitem}
\usepackage{caption}
\usepackage{multicol}
\begin{document}
\definecolor{darkgreen}{rgb}{0,0.5,0}
\definecolor{orange}{rgb}{1,0.5,0}
\definecolor{teal}{rgb}{0,0.5,0.5}
\definecolor{darkpurple}{rgb}{0.5, 0, 0.5}



\title{Generating Automatic Feedback on UI Mockups with Large Language Models}

\author{Peitong Duan}
\email{peitongd@berkeley.edu}
\affiliation{%
  \institution{UC Berkeley}
  \city{Berkeley}
  \country{CA, USA}
}

\author{Jeremy Warner}
\email{jeremy.warner@berkeley.edu}
\affiliation{%
  \institution{UC Berkeley}
  \city{Berkeley}
  \country{CA, USA}
}

\author{Yang Li}
\email{liyang@google.com}
\affiliation{%
  \institution{Google Research}
  \city{Mountain View}
  \country{CA, USA}
}

\author{Bjoern Hartmann}
\email{bjoern@eecs.berkeley.edu}
\affiliation{%
  \institution{UC Berkeley}
  \city{Berkeley}
  \country{CA, USA}
}
\begin{abstract}
Feedback on user interface (UI) mockups is crucial in design. However, human feedback is not always readily available. We explore the potential of using large language models for automatic feedback. Specifically, we focus on \changes{applying GPT-4 to automate heuristic evaluation}, which currently entails a human expert assessing a UI’s compliance with a set of design guidelines. We implemented a Figma plugin that takes in a UI design and a set of written heuristics, and renders automatically-generated feedback as constructive suggestions. We assessed performance on 51 UIs using three sets of guidelines, compared GPT-4-generated design suggestions with those from human experts, and conducted a study with 12 expert designers to understand fit with existing practice. We found that GPT-4-based feedback is useful for catching subtle errors, improving text, and considering UI semantics, but feedback also decreased in utility over iterations. Participants described several uses for this plugin despite its imperfect suggestions.
\end{abstract}



\begin{teaserfigure}
  \includegraphics[width=\columnwidth]{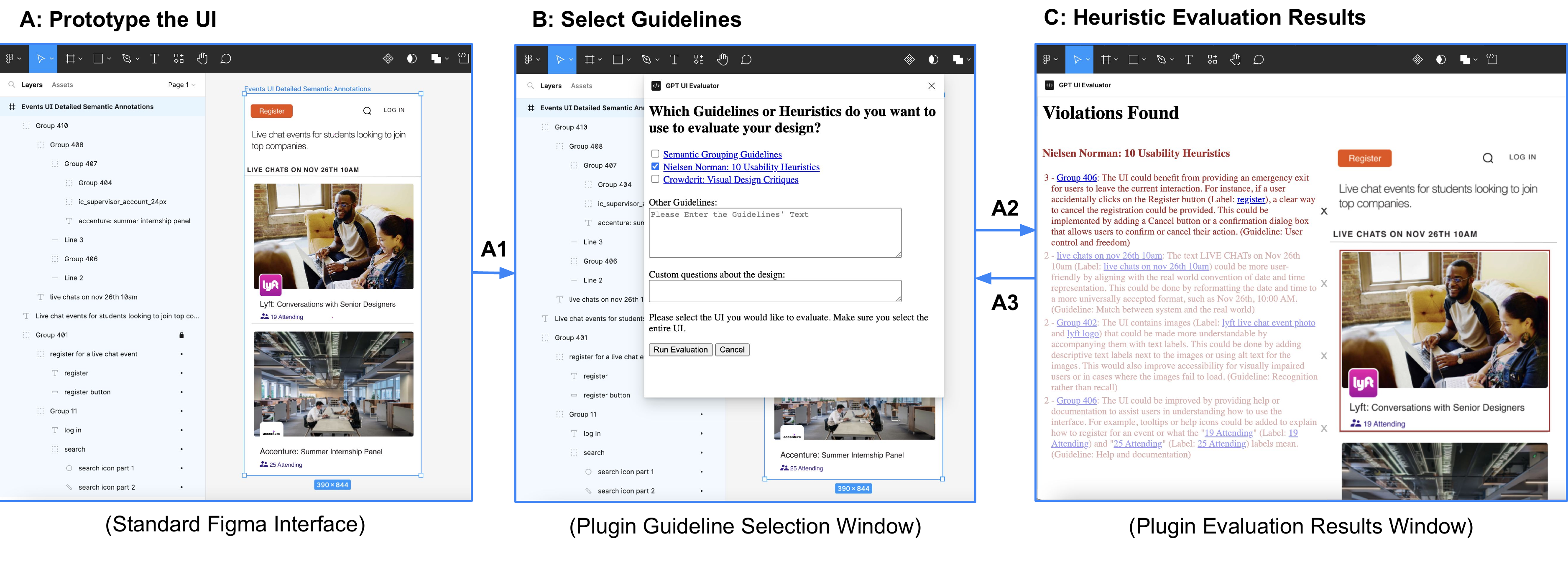}
  \caption{Diagram illustrating the UI prototyping workflow using this plugin.
  First, the designer prototypes the UI in Figma (Box A) and then runs the plugin (Arrow A1).
  The designer then selects the guidelines to use for evaluation (Box B) and runs the evaluation with the selected guidelines (Arrow A2).
  The plugin obtains evaluation results from the LLM and renders them in an interpretable format (Box C).
The designer uses these results to update their design and reruns the evaluation (Arrow A3).
  The designer iteratively revises their Figma UI mockup, following this process, until they have achieved the desired result.}
  \label{fig:teaser}
  \Description{A flowchart with three boxes. The first Box contains a screenshot of Figma with a UI mockup. There is an arrow labeled “A1” pointing from the first Box to the second Box. The second Box contains this Figma screen with the plugin window open, which contains checkboxes for common guidelines, like NN Usability Heuristics, and a large text field to enter in custom guidelines. There is an arrow “A2” pointing from the second box to the third box and an arrow “A3” pointing from the third box to the second box. The third box contains a plugin screen with the evaluation results. There a list of violation explanation text on the left and the UI screenshot on the right.}
\end{teaserfigure}


\maketitle

\section{Introduction}
User interface (UI) design is an essential domain that shapes how humans interact with technology and digital information. Designing user interfaces commonly involves iterative rounds of feedback and revision. Feedback is essential for guiding designers towards improving their UIs. While this feedback traditionally comes from humans (via user studies and expert evaluations), recent advances in computational UI design enable automated feedback. However, automated feedback is often limited in scope (e.g., the metric could only evaluate layout complexity) and can be challenging to interpret \cite{schoop2022predicting}. While human feedback is more informative, it is not readily available and requires time and resources for recruiting and compensating participants.

One method of evaluation that still relies on human participants today is \textit{heuristic evaluation}, where an experienced evaluator checks an interface against a list of usability heuristics (rules of thumb) developed over time, such as Nielsen's 10 Usability Heuristics~\cite{nielsen10heuristics}. Despite appearing straightforward, heuristic evaluation is challenging and subjective \cite{Nielsen1990HeuristicEO}, dependent on the evaluator's previous training and personality-related factors \cite{LING2009382}. These limitations further suggest an opportunity for AI-assisted evaluation.

There are several reasons why LLMs could be suitable for automating heuristic evaluation. The evaluation process primarily involves rule-based reasoning, which LLMs have shown capacity for \cite{openai2023gpt4}. Moreover, design guidelines are predominately in text form, making them amenable for LLMs, and the language model could also return its feedback as text-based explanations that designers prefer~\cite{lee2020guicomp}. Finally, LLMs have demonstrated the ability to understand and reason with mobile UIs \cite{10.1145/3544548.3580895}, as well as generalize to new tasks and data \cite{Lu_Grover_Abbeel_Mordatch_2022, sanh2022multitask}. However, there are also reasons that suggest caution for using LLMs for this task. For one, LLMs only accept text as input, while user interfaces are complex artifacts that combine text, images, and UI components into hierarchical layouts. In addition, LLMs have been shown to hallucinate \cite{lee2023benefits} (i.e., generate false information) and may potentially identify incorrect guideline violations. This paper explores the potential of using LLMs to carry out heuristic evaluation automatically. In particular, we aim to determine their performance, strengths and limitations, and how an LLM-based tool can fit into existing design practices.

To explore the potential of LLMs in conducting heuristic evaluation, we built a tool that enables designers to run automatic evaluations on UI mockups and receive text-based feedback. We package this system as a plugin for Figma~\cite{figma}, a popular UI design tool. Figure \ref{fig:teaser} illustrates the iterative usage of this plugin. The designer prototypes their UI in Figma, and then selects a set of guidelines they would like to use for evaluation in our plugin. The plugin returns the feedback, which the designer uses to revise their mockup. The designer can then repeat this process on their edited mockup. To improve the LLM's performance and adapt to individual preferences, designers can provide feedback on each generated suggestion, which is integrated into the model for the next round of evaluation. The plugin produces UI mockup feedback by querying an LLM with the guidelines' text and a JSON representation of the UI. The LLM then returns a set of detected guideline violations. Instead of directly stating the violations, they are phrased as constructive suggestions for improving the UI. As LLMs can only process text and have a limited context window, we developed a JSON representation of the UI that concisely captures the layout hierarchy and contains both semantic (text, semantic label, element type) and visual (location, size, and color) details of each element and group in the UI. \changes{To further accommodate context window limits, we scoped the plugin to evaluate only static (i.e., non-interactive) UI mockups, one screen at a time.}

We conducted an exploration of how several current state of the art LLMs perform on this task and found that GPT-4 had the best performance by far. Hence, we solely focus on GPT-4 for the remaining studies. To assess GPT-4's performance in conducting heuristic evaluation on a large scale, we carried out a study where three design experts rated the accuracy and helpfulness of its heuristic evaluation feedback for 51 distinct UIs. To compare GPT-4's output with feedback provided by human experts, we conducted a heuristic evaluation study with 12 design experts, who manually identified guideline violations in a set of 12 UIs. Finally, to qualitatively determine GPT-4's strengths and limitations and its performance as an iterative design tool, we conducted a study with another group of 12 design experts, who each used this tool to iteratively refine a set of 3 UIs and evaluated the LLM feedback each round. For all three studies, we used diverse guidelines covering visual design, usability, and semantic organization to generate design feedback.

We found that GPT-4 was generally accurate and helpful in identifying issues in poor UI designs, but its performance became worse after iterations of edits that improved the design, making it unsuitable as an iterative tool. Furthermore, its performance varied, depending on the guideline. GPT-4 generally performed well on straightforward checks with the data available in the UI JSON and worse when the JSON differed from what was visually or semantically depicted in the UI. Finally, although GPT-4's feedback is sometimes inaccurate, most study participants still found this tool useful for their own design practices, as it was able to catch subtle errors, improve the UI’s text, and reason with the UI's semantics. They stated that the errors made by GPT-4 are not dangerous, as there is a human in the loop to catch them, and suggested various use cases for the tool. Finally, we also distilled a set of concrete limitations of GPT-4 for this task.

In summary, even with today's limitations, GPT-4 can already be used to automatically evaluate some heuristics for UI design; other heuristics may require more visual information or other technical advancements. However, designers accepted occasional imperfect suggestions and appreciated GPT-4's attention to detail. This implies that while LLM tools will not replace human heuristic evaluation, they may nevertheless soon find a place in design practice.

Our contributions are as follows:
\begin{itemize}
    \item A Figma plugin that uses GPT-4 to automate heuristic evaluation of UI mockups with arbitrary design guidelines.
    \item An investigation of GPT-4's capability to automate heuristic evaluations through a study where three human participants rated the accuracy and helpfulness of LLM-generated design suggestions for 51 UIs. 
    \item A comparison of the violations found by this tool with those identified by human experts.
    \item An exploration of how such a tool can fit into existing design practice via a study where 12 design experts used this tool to iteratively refine UIs, assessed the LLM-generated feedback, and discussed their experiences working with the plugin.
\end{itemize}

\section{Related Work}

\subsection{AI-Enhanced Design Tools}
Before the widespread use of generative AI, research in AI-enhanced design tools explored a variety of model architectures to accomplish a wide range of tasks. For instance, Lee et al. built a prototyping assistance tool (GUIComp) that provides multi-faceted feedback for various stages of the prototyping process. GUIComp uses an auto-encoder to support querying UI examples for design inspiration and separate convolutional neural networks to evaluate the visual complexity of the UI prototypes and predict salient regions \cite{lee2020guicomp}. Other studies have utilized computer vision techniques to predict saliency in graphical designs \cite{10.1145/3379337.3415825} and perceived tappability \cite{schoop2022predicting, 10.1145/3290605.3300305}. Deep learning models have been developed for generation \cite{cheng2023play}, autocompletion \cite{10.1145/3490034}, and optimization \cite{10.1145/3313831.3376589, 10.1145/3313831.3376593, 10.1145/2901790.2901817} of UI layouts. One limitation of these techniques is that a separate model is needed for each type of task. In addition, study participants had difficulty interpreting the feedback from these models~\cite{schoop2022predicting} and would have liked natural language explanations of detected design issues~\cite{lee2020guicomp}. Our work addresses both of these limitations. First, our system supports arbitrary guidelines that evaluate various aspects of the UI design as input. Furthermore, the language model uses natural language to explain each detected guideline violation. 

\subsection{Applications of Generative AI in Design}
The recent emergence of generative AI, such as GPT, has led to various applications in design support. 
Park et al. carried out two studies that employ LLMs to simulate user personas in online social settings. They used GPT-3 to generate interactions on social media platforms
as testing data for these platforms \cite{10.1145/3526113.3545616}. They later expanded on this work to build agents that could remember, reflect on, and retrieve memories from interacting with other agents to realistically simulate large-scale social interactions \cite{park2023generative}. H\"am\"al\"ainen et al. used GPT-3 to generate synthetic human-like responses to survey questionnaires about video game experiences \cite{10.1145/3544548.3580688}. 
Finally, Wang et al. investigated the feasibility of using LLMs to interact with UIs via natural language \cite{10.1145/3544548.3580895}. They developed prompting techniques for tasks like screen summarization, answering questions about the screen, generating questions about the screen, and mapping instructions to UI actions. 
Researchers have also begun to create design tools that use Generative AI. Lawton et al. built a system where a human and generative AI model collaborate in drawing, and ran an exploratory study on the capabilities of this tool \cite{10.1145/3563657.3595977}. Stylette allows users to specify design goals in natural language and uses GPT to infer relevant CSS properties \cite{10.1145/3491102.3501931}. Perhaps most similar to our work is a study by Petridis et al. \cite{10.1145/3544549.3585628}, who explored using LLM prompting in creating functional LLM-based UI prototypes. Their study findings showed that LLM prompts sped up prototype creation and clarified LLM-based UI requirements, which led to the development of a Figma Plugin for automated content generation and determination of optimal frame changes. These existing studies, however, have not examined the application of LLMs as a general-purpose evaluator for mobile UIs of any category with a diverse set of heuristics. 

\subsection{AI-enhanced Software Testing}
Another domain of UI evaluation is testing the functionality of the GUI (i.e., ``software testing''). Existing LLM-based approaches include Liu et al.'s method \cite{liu2023chatting}, which uses GPT-3 to simulate a human tester that would interact with the GUI.
Their system had greater coverage and found more bugs than existing baselines, and also identified new bugs on Google Play Store apps. 
Wang et al. conducted a comprehensive literature review on using LLMs for software testing. They analyzed various studies that used LLMs for unit test generation, validation of test outputs, test input generation, analyzing bugs, fixing identified bugs in code, and identifying and correcting bugs.
Contrary to software testing, our study focuses on evaluating GUI mockups, which is at an earlier stage of the UI development process. 
Furthermore, evaluation of mockups and software are intrinsically different; mockup evaluation focuses on adherence to design guidelines and user feedback, whereas software testing focuses on finding bugs in the implementation. 
Prior to LLMs, Chen et al. utilized computer vision techniques to identify discrepancies between the UI mockup and implementation \cite{chen2017xray}. Their system could identify differences in positioning, color, and size of corresponding elements. However, their evaluation requires a UI mockup as the benchmark, while our system could carry out  evaluation using any set of heuristics. 

\subsection{Heuristics and Design Guidelines}
An essential aspect of the design process is gathering feedback to improve future iterations.
One central way designers generate feedback is to conduct \textit{heuristic evaluations}~\cite{Nielsen1990HeuristicEO, 10.1145/142750.142834}, which uses a set of guidelines to identify and characterize undesired interface characteristics as violations of specific guidelines.
While initially designed for desktop interfaces, other work has adapted heuristic evaluation to more devices and domains \cite{10.1145/642611.642642, 10.1145/1357054.1357282, Mi2014AHC, 10.1145/2987592.2987617}.
In general, researchers have developed design guidelines for a vast category of devices, tasks, and populations including accessible data visualizations \cite{elavsky2022accessible}, multi-modal touchscreen graphics \cite{10.1145/3403933}, smart televisions \cite{10.1145/2932206.2932212}, ambient lighting interactions \cite{10.1145/2836041.2836069}, hands-free speech interaction \cite{10.1145/3236112.3236149}, navigation in virtual environments \cite{10.1145/302979.303062}, website readability \cite{10.1145/3064663.3064711}, supporting web design for aging communities \cite{10.1145/1056808.1057050, 10.1145/1090785.1090810}, and for cross-cultural design considerations \cite{10.1145/3058555.3058574}.
A widely-used set of guidelines is Nielsen's 10 Usability Heuristics~\cite{nielsen10heuristics}, a set of general principles for interaction design. Luther et al. surveyed design textbooks and other resources and compiled a comprehensive set of specific critique statements for the visual design of an interface, which were organized into 7 visual design principles~\cite{10.1145/2556420.2556788}. 
Recently, Duan et al. developed a set of 5 specific and actionable guidelines for organizing UI elements based on their semantics (i.e., functionality, content, or purpose) to help design clear and intuitive interfaces~\cite{duan2023towards}.
While these guidelines are meant to encode common design patterns and errors distilled from design expert guidance, they still require a human to interpret and apply them, making adapting to a new set of guidelines time-consuming, especially for novice designers.  
Our work builds off of these design guidelines as a means of focusing and justifying the LLM's design suggestions and feedback.

\subsection{User Interfaces for Design Feedback}
Prior research has explored several ways to support designers as they both give and receive feedback across a range of media \cite{10.1145/2724660.2724670, luther2015structuring, 10.1145/3581641.3584035, 10.1145/2642918.2647390, 10.1145/2984511.2984552}.
Cheng et al. explore the process of publicly gathering design feedback from online forums \cite{Cheng2020CritiqueM} and list several design considerations for feedback systems.
For supporting in-context feedback for graphic designs, CritiqueKit \cite{Fraser2017CritiqueKitAM} showcased a UI for providing and improving real-time design feedback, while Charrette \cite{TranOLeary2018CharretteSI} supported organizing and sharing feedback on longer histories and variations of a design.
A study by Ngoon et al. showcased reusing expert feedback suggestions and adaptive guidance as two ways of improving creative feedback by making the feedback more specific, justified, and actionable \cite{Ngoon2018InteractiveGT}.
This notion of adaptive conceptual guidance is further explored by Shöwn \cite{Ngoon2021ShwnAC}, demonstrating the utility of adapting presented design suggestions and examples automatically given the user's current working context. 
Our plugin provides in-context design feedback grounded by this prior work on user interfaces for design feedback, while automatically generating the provided feedback and design suggestions.

\section{System Details}
In this section, we describe the set of design goals for an automatic LLM-driven heuristic evaluation tool, how they are realized in our system, the underlying implementation, techniques to improve the LLM's performance, and explorations of alternative prompt designs and various LLM models for this task.

\subsection{Design Goals}
Based on design principles and expected LLM behavior, we came up with a set of goals that lay out what an automatic LLM-based heuristic evaluation tool should be able to do. The goals are as follows:
\begin{enumerate}[labelindent=0pt]    
    \item The tool should be able to accommodate arbitrary UI prototypes; designers should be able to use this tool to perform heuristic evaluations on their mockups and identify potential issues, before implementation.
    \item The tool should be heuristic-agnostic, so different guidelines or heuristics can be used.
    \item The guideline violations detected by the LLM should be presented in a way that adheres to the principles of effective feedback \cite{sadler1989formative}.
    \item The LLM generated feedback should be presented in the context of the critiqued design. This is to narrow the gulf of evaluation, making it easier for designers to interpret the feedback.
    \item Finally, in case the LLM makes a mistake, the designer should be able to hide feedback they find unhelpful. This data should also be sent to the LLM to improve its prediction accuracy.
\end{enumerate}
\subsection{Design Walkthrough}
\begin{figure*}
  \centering
  \includegraphics[width=\textwidth]{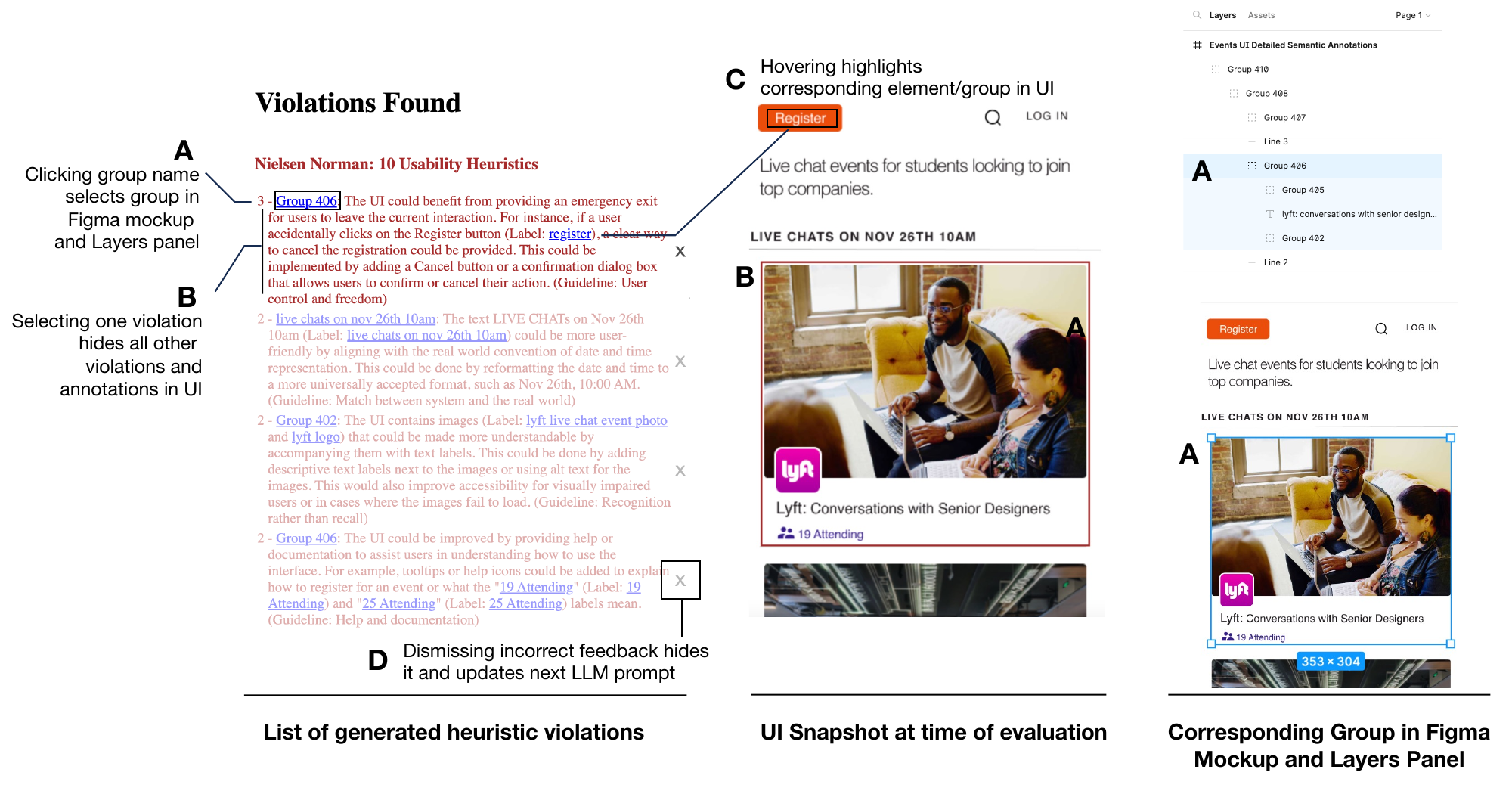}
  \caption{Illustration of plugin interactions that contextualize text feedback with the UI. ``A'' shows that clicking on a link in the violation text selects the corresponding group or element in the Figma mockup and Layers panel. ``B'' shows the ``click to focus'' feature, where clicking on a violation fades the other violations and draws a box around the corresponding group in the UI screenshot. ``C'' illustrates that hovering over a group or element link draws a blue box around the corresponding element in the screenshot. ``D'' points out that clicking on the `X' icon of a violation hides it and adds this feedback to the LLM prompt for the next round of evaluation.}
    \label{fig:plugininteractions}
  \Description{Three side by side images. A screenshot of the violations text, a screenshot of a UI mockup, and a screenshot of the layers panel with a screenshot of the Figma mockup below. A box is drawn around ‘Group 406’ and it corresponds to the group selected in the layers panel and the Figma mockup.}
\end{figure*}
We built our tool as a plugin for Figma, enabling designers to evaluate any Figma mockup (Goal 1).
%
Figure \ref{fig:teaser} illustrates this plugin's step-by-step usage with interface screenshots.
The designer first prototypes their UI in Figma and runs the plugin (Figure \ref{fig:teaser} Box A). Due to context window limitations, the plugin only evaluates a single UI screen at a time. \changes{Furthermore, it only assesses static mockups, as evaluation of interactive mockups may require multiple screens as input or more complex UI representations, which could exceed the LLM's context limit.} After starting the plugin, it opens up a page to select guidelines to use for heuristic evaluation (Box B). Designers can select from a set of well-known guidelines, like Nielsen's 10 Usability Heuristics, or enter any list of heuristics they would like to use (Goal 2). They can also select more than one set of guidelines for the evaluation.

Once the LLM completes the heuristic evaluation, text explanations of all violations found and a UI screenshot are rendered back to the designer (Figure \ref{fig:teaser} Box C). This ``UI Snapshot'' serves as a reference to the state of the mockup at the time of evaluation, in case the designer makes any changes based on the evaluation results. Each violation explanation contains the name of the violated guideline and is phrased as constructive feedback, following the guidelines set by Sadler et al. \cite{sadler1989formative} (Goal 3). According to Sadler, effective feedback is specific and relevant, highlighting the performance gap and providing actionable guidance for improvement. To accomplish this, the feedback must include these three things: 1) the expected standard, 2) the gap between the quality of work and the standard, and 3) what needs to be done to close this gap. Our design feedback adheres to Sadler's principles and starts by stating the standard set by the guideline, followed by the issue with the current design (the gap between the design and expected standard), and concludes with advice on fixing the issue. Figure \ref{fig:goodandbadsuggestions} provides four examples of these explanations. 

The plugin also includes several features that help designers contextualize the text feedback with corresponding UI elements. 
(Goal 4). Figure \ref{fig:plugininteractions} illustrates these features. 
Selecting a violation fades the other suggestions and draws a box around the relevant group or element in the screenshot, as shown in Figure \ref{fig:plugininteractions} (B). In addition, all UI elements and groups mentioned are rendered as links. Hovering over a link draws a box over the corresponding group or element in the screenshot (C), and clicking on the link selects the item in the Figma mockup and Layers panel (A), streamlining the editing process. Finally, to address Goal 5, if the designer finds a suggestion incorrect or unhelpful, they can click on its `X' icon to hide it (D). Hiding the violation sends feedback to the LLM for subsequent evaluation rounds so this violation will not be shown again.

After the designer revises their mockup based on LLM feedback, they can rerun the evaluation to generate new suggestions. 
This usage is intended to match the iterative feedback and revision process during design.
The plugin uses the information from the Layers panel of Figma to create the text-based representation of the mockup (discussed in more detail in the next section). Hence, it relies on accurate names for groups in the Layers panel to convey semantic information about the UI; for instance, the group containing icons in the navbar should be named ``navbar''. Designers must manually add these names, so they are often missing. To address this, we implemented an auxiliary label generation feature that can be run before evaluation to generate group names automatically (based on their contents).

\subsection{Implementation}
\begin{figure*}
  \centering
  \includegraphics[width=\textwidth]{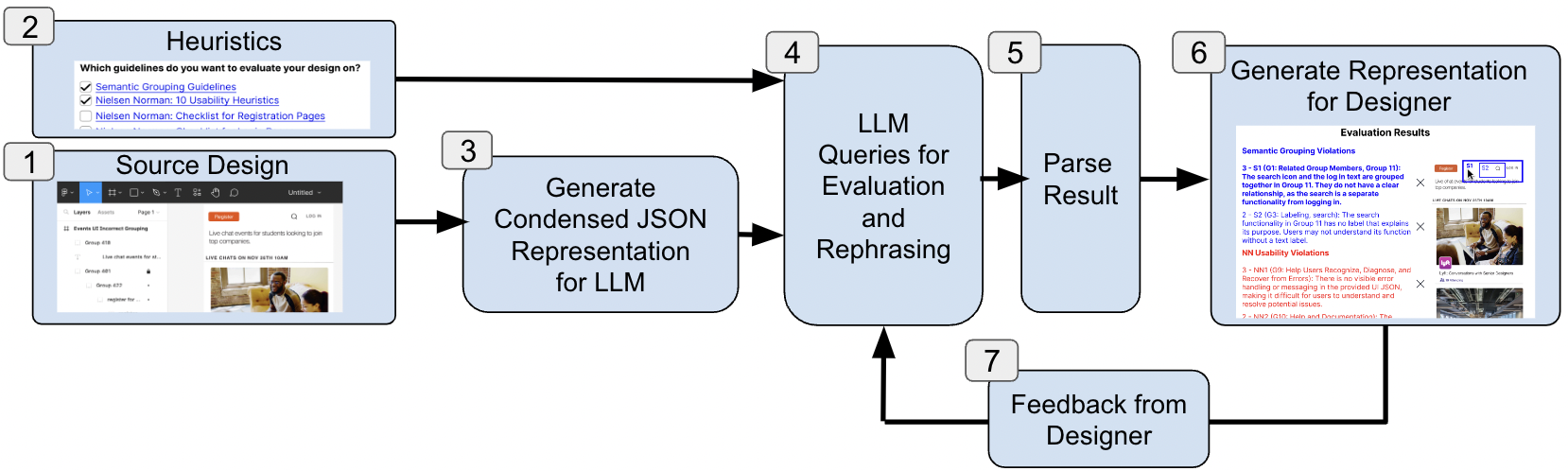}
  \caption{
  Our LLM-based plugin system architecture. The designer prototypes a UI in Figma (Box 1), and the plugin generates a UI representation to send to an LLM (3). The designer also selects heuristics/guidelines to use for evaluating the prototype (2), and a prompt containing the UI representation (in JSON) and guidelines is created and sent to the LLM (4). After identifying all the guideline violations, another LLM query is made to rephrase the guideline violations into constructive design advice (4). The LLM response is then programmatically parsed (5), and the plugin produces an interpretable representation of the response to display (6). The designer dismisses incorrect suggestions, which are incorporated in the LLM prompt for the next round of evaluation, if there is room in the context window (7). 
  }
    \label{fig:systemdiagram}
\Description{A flow chart illustrating the system design of the plugin. Source design is labeled 1 and Heuristics is labeled 2. There is an arrow from Source Design to Generate Condensed JSON Representation for LLM (labeled 3). Generate Condensed JSON Representation for LLM and Heuristics both point to LLM Queries for Evaluation and Rephrasing  (labeled 4), which points to Parse Result (5), which points to Generate Representation for Designer (6). Generate Representation for Designer points to LLLM Queries for Evaluation and Rephrasing going through a box label 7 saying Feedback from Designer}
\end{figure*}
\begin{figure}
  \centering
  \includegraphics[width=\columnwidth]{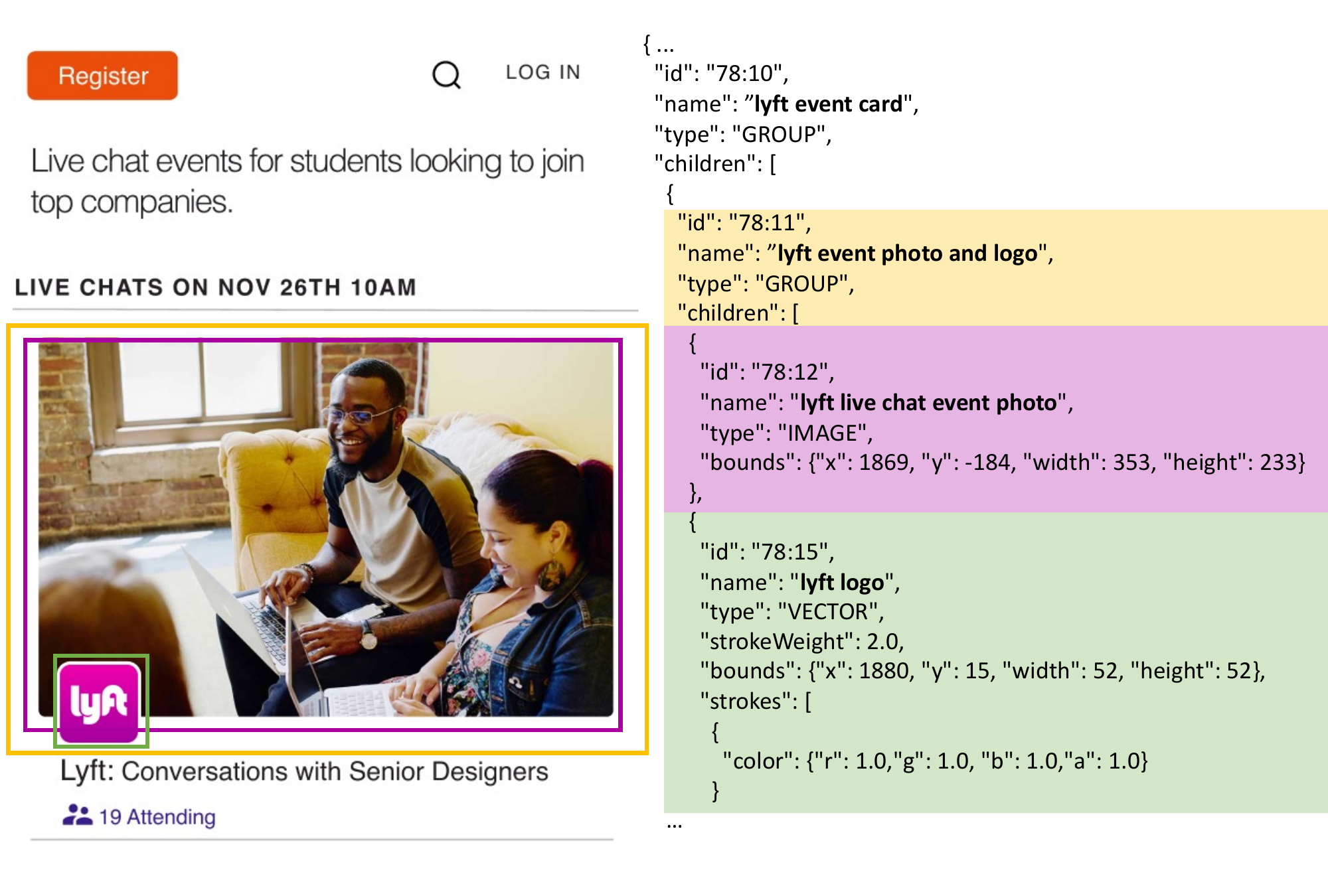}
  \caption{An example portion of a UI JSON. It has a tree structure, where each node has a list of child nodes (the ``children'' field). Each node in this JSON is color-coded with its corresponding group or element in the UI screenshot. The node named ``lyft event photo and logo'' is a group (``type: GROUP'') consisting of a photo of the live chat event (``lyft live chat event photo'') and the Lyft logo (``lyft logo''). The JSON node for the photo contains its location information (``bounds''), type (``IMAGE''), and unique identifier (``id''). The JSON node for ``lyft logo'' contains its location and some stylistic information, like the stroke color and stroke weight for its white border. 
  }
    \label{fig:jsonexample}
\Description{UI Mockup Screenshot on the left and a snippet of JSON on the right. There are boxes marking corresponding sections in the JSON and screenshot. There is a box around the photo and the section is highlighted in JSON, which contains the photo name, type (image), and bounding box coordinates. Similar correspondences are drawn between the Lyft logo and the group.}
\end{figure}
We implemented this plugin in Typescript using the Figma Plugin API. The plugin makes an API request to OpenAI's GPT-4 for LLM queries. Since LLMs can only accept text as input, the plugin takes in a JSON representation of the UI. While multi-modal models exist that could take in both the UI screenshot and guidelines text (e.g., \cite{liu2023visual}), we found that its performance was considerably worse than GPT-4's for this task (at the time).

Our JSON format captures the DOM (Document Object Model) structure of the UI mockup, and is similar in structure and content to the HTML-based representation used by \cite{10.1145/3544548.3580895} that performed well on UI-related tasks. Figure \ref{fig:jsonexample} contains an example portion of a UI JSON with corresponding groups and elements marked in the UI screenshot. The tree structure is informative of the overall organization of the UI, with UI elements (buttons, icons, etc.) as leaves and groups (of elements and/or smaller groups) as intermediate nodes. Each node in this JSON tree contains semantic information (text labels, element or group names, and element type) and visual data (x,y-position of the top left corner, height, width, color, opacity, background color, font, etc.) of its element or group. Hence, this JSON representation captures both semantic and visual features of the UI, which supports the evaluation of various aspects of the design and differentiates it from the representation used by \cite{10.1145/3544548.3580895} that captures only semantic information. This JSON representation is constructed from the data (e.g., group/element names) and grouping structure found in the Layers panel of Figma, which are editable by designers.

Figure \ref{fig:systemdiagram} shows the core system design of the plugin. Due to the context window limits of GPT-4, we remove all unnecessary or redundant information and condense verbose details into a concise JSON structure (Box 3). This condensed JSON representation and guideline text are combined into a prompt sent to the LLM. After the LLM returns the identified guideline violations, another query is sent to the LLM to convert these violations into constructive advice (Box 4). This chain of prompts is illustrated in Figure \ref{fig:promptdiagram} (Appendix), which describes the components of each prompt. The LLM response is parsed by the TypeScript code (Box 5) and rendered into an interpretable format for designers (Box 6). Figma IDs for each element and group are stored internally, which supports selection of elements or groups in the mockup via links (Figure \ref{fig:plugininteractions}, A) and quick access to their layout information. Layout information is used to draw boxes around elements and groups in the screenshot, as shown in Figure \ref{fig:plugininteractions} (B and C). Finally, unhelpful suggestions that were dismissed by the designer are incorporated into the prompt for the next round of evaluation (Box 7), if there is room in the context window. 
The label generation feature is also executed via an LLM call, with the prompt containing JSON data of all unnamed groups and instructions for the LLM to create a descriptive label for each JSON based on its contents. 

\subsection{Improving LLM Performance}
We chose the most advanced GPT version available (GPT-4), as it has the strongest reasoning abilities~\cite{openai2023gpt4}. However, GPT-4 does not support fine-tuning and has a context window limit of 8.1k tokens. This context window limit leaves inadequate room for few-shot and ``chain-of-thought''~\cite{wei2023chainofthought} examples because each Figma UI JSON requires around 3-5k tokens, the guidelines text take up to 2k tokens, and few-shot and chain-of-thought examples both require the corresponding UI JSONs. Due to these limitations, our method for improving GPT-4's performance entailed adding explicit instructions in the prompt to avoid common mistakes, as shown in Figure \ref{fig:promptdiagram} (Appendix). 
Finally, we set the temperature to 0 to ensure GPT-4 returns the most probable violations. 

The remaining space in the context window was allocated to suggestions that were dismissed (hidden) by designers. Incorporating this feedback targets areas of poor performance specific to the UI being evaluated and also adapts GPT-4's feedback to the designer's preferences. Since the UI JSON is already provided in the prompt, this feedback does not require much space. However, the UI may have changed due to edits, so JSONs of the groups/elements for a dismissed violation are still included, but they are considerably smaller than the entire UI JSON. These items are incorporated in the conversation history of the next prompt, as examples of inaccurate suggestions (see Figure \ref{fig:promptdiagram} in the Appendix). In addition, we ask GPT-4 to reflect on why it was wrong and add this prompt and its response to the conversation history. This ``self-reflection'' has been shown to improve LLM performance~ \cite{shinn2023reflexion}.

\subsection{Exploration of Alternative Prompt Compositions}
\begin{table}

\centering
\begin{tabular}{|p{3cm}|p{1.5cm}|p{1.5cm}|}
\hline
\textbf{Prompt Condition} & \textbf{Total \newline Violations} & \textbf{Helpful Violations}\\ \hline
Complete (Plugin) & 63 & 38\\ \hline
One Call & 62 & 31\\ \hline
No Heuristics & 50 & 14\\ \hline
General UI Feedback & 57 & 24\\ \hline
\end{tabular}
\vspace{-1.5em}
\caption*{}
\begin{tabular}{|p{3cm}|p{1.5cm}|p{1.5cm}|}
\hline
\textbf{LLM} & \textbf{Total \newline Violations} & \textbf{Helpful Violations}\\ \hline
GPT-4 (Plugin) & 63 & 38\\ \hline
GPT-3.5-16k & 228 & 23\\ \hline
Claude 2 & 7 & 1\\ \hline
PaLM 2 & 12 & 3\\ \hline
\end{tabular}
\vspace{0.5em}
\caption{The top table compares the total number of violations and the number of helpful violations (based on the authors' judgement) found in 12 UI mockups for different prompt compositions. The ``Complete (Plugin)'' condition refers to the prompt composition used in the plugin. The bottom table compares the total number of violations and the number of helpful violations (based on the authors' judgement) found in the 12 UI mockups by each LLM, with GPT-4 being used in the plugin.}
\end{table}
\label{tab:promptablation}
We investigate how different prompt components influence GPT-4's output to identify potential opportunities for simplifying our complex prompt. For our analysis, we used 12 distinct mockups of mobile UIs taken from the Figma community. Furthermore, we used three sets of heuristics covering different aspects of UI design: Nielsen's 10 Usability Heuristics~\cite{10.1145/142750.142834}, Luther et al.'s visual design principles compiled in  ``CrowdCrit'' \cite{10.1145/2556420.2556788}, and Duan et al.'s 5 semantic grouping guidelines~\cite{duan2023towards}. These 12 UIs and three sets of heuristics were consistently used in all subsequent analyses and studies in this paper, except for the Performance Study, which used a larger set of 51 UIs. 
We query the LLM with prompt variations and then compute the total number of reported violations and the number of helpful violations (based on the authors' judgment), and we also qualitatively examine the violations. We consider a violation to be helpful if it is both accurate and would lead to an improvement in the design. Table \ref{tab:promptablation} (``Prompt Condition'') compares violation counts for each condition with the complete prompt chain.
\subsubsection{One Call}
Our prompt chain makes two LLM calls -- one to carry out the heuristic evaluation and the other to rephrase results into constructive feedback (Appendix Figure \ref{fig:promptdiagram}). We examine the effects of combining these two into a single call, as this would reduce latency. 
Quantitatively, the total number of violations remained similar, but the number of helpful violations was lower. However, more importantly, the output was never formatted correctly with one call, and the format also varied across different calls. 
Furthermore, GPT-4 sometimes omitted other important details, such as how to fix the violation. 
Since correct output formatting is necessary for the plugin to parse and render the violations, combining the two calls is not feasible.

\subsubsection{No Heuristics}
The detailed heuristics text occupies a lot of space in the LLM's context window, so we examined the performance without including them in the prompt. We edited prompts to look for ``visual design issues'', ``usability issues'', or ``semantic group issues'' instead of passing in the heuristics.

%
Table \ref{tab:promptablation} (top) shows that GPT-4 provided fewer suggestions total (50 vs. 63) and considerably fewer helpful suggestions when heuristics were not included in the prompt (14 vs. 38). 
Qualitatively, the suggestions for Crowdcrit and Nielsen were similar to those from the complete prompt, though the suggestions were more thorough when the Crowdcrit heuristics were included.
However, for Semantic Grouping, not passing in the heuristics resulted in only violations that concerned the semantic relatedness of group members, whereas passing in the guidelines resulted in a more diverse set of issues found. We conclude that while the LLM could give plausible UI feedback without passing in heuristics, the quality of the suggestions is worse.

\subsubsection{General UI Feedback}
Finally, we investigate how GPT-4 responds without specific guidance when prompted for general UI feedback. 
We removed all mentions of ``guidelines''  in the prompt and replaced ``violations'' with ``feedback.'' Quantitatively, the performance for this condition was worse. Qualitatively, GPT-4 still carried out heuristic evaluation to an extent, as the issues were grounded in existing design conventions, but in a less rigorous and organized manner. 
Compared to the complete prompt, the feedback was less diverse, and the LLM often focused on only one type of issue (e.g., misalignment) when there were other types of violations. 
We conclude that GPT-4 can produce plausible output when asked for general UI feedback, but specific guidance produces higher quality and more diverse suggestions.

\subsection{Comparison with other LLMs}
We explored the potential of other state-of-the-art LLMs in carrying out this task: Claude 2, GPT-3.5-turbo-16k, and PaLM 2. Llama 2 was considered but excluded because its 4k context window size is insufficient for the task. Similar to the prompt analysis, we compute the total number of violations found and the number of helpful violations. 

We found that Claude 2, GPT-3.5-turbo-16k, and PaLM 2 all had considerably worse performance than GPT-4, as shown in Table \ref{tab:promptablation} (bottom). Claude 2 and PaLM 2 found very few violations; Claude 2 only found violations in 4 UIs, and PaLM 2 only identified one violation per UI, even after adjusting the prompt to indicate more than one violation per UI. In fact, all 12 UIs have multiple violations, as later confirmed in a heuristic evaluation by human experts. The few violations found by these two LLMs were mostly unhelpful, such as suggesting the dollar sign needs a text label. 
GPT-3.5-turbo-16k had the opposite behavior, finding nearly 4 times as many violations as GPT-4. However, most of the time, it indiscriminately applied the same guideline to every element of the appropriate type, regardless if there is an issue (e.g., stating the font is difficult to read for every text element). 
This behavior also meant that most of its helpful violations were found by chance, despite finding fewer helpful violations than GPT-4. 
Finally, GPT-3.5-turbo-16k and PaLM 2 had difficulty following the prompt's instructions, often formatting the output incorrectly (with a separate rephrasing call) or making the mistakes they were told to avoid, such as returning violations regarding the mobile status bar.

These models are all smaller than GPT-4, with billions of parameters, compared to GPT-4's 1.7 trillion \cite{openai2023gpt4}. These models have also been shown to have worse reasoning skills \cite{unknown}. These factors likely contributed to their poor performance in this task. Since GPT-4 has the best performance by far, we solely focus on GPT-4 for the remaining three studies on the plugin.

\section{Study Method}
\begin{figure}
  \centering
\includegraphics[width=\columnwidth]{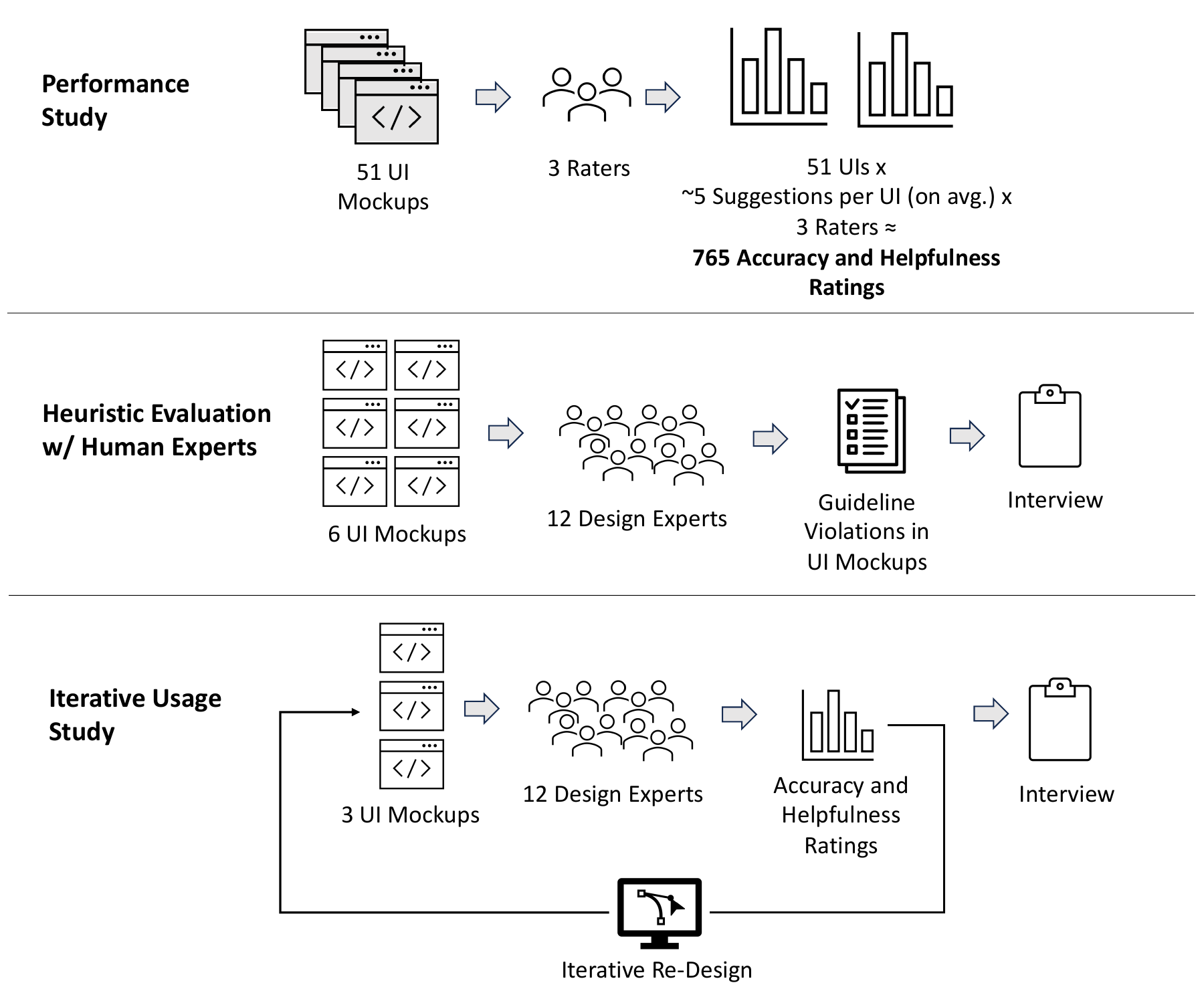}
  \caption{
 An illustration of the formats of the three studies. The Performance Study consists of 3 raters evaluating the accuracy and helpfulness of GPT-4-generated suggestions for 51 UI mockups. The Heuristic Evaluation Study with Human Experts consists of 12 design experts, who each looked for guideline violations in 6 UIs, and finishes with an interview asking them to compare their violations with those found by the LLM. Finally, the Iterative Usage study comprises of another group of 12 design experts, each working with 3 UI mockups. For each mockup, the expert iteratively revises the design based on the LLM's valid suggestions and rates the LLM's feedback, going through 2-3 rounds of this per UI. The Usage study concludes with an interview about the expert's experience with the tool.}
    \label{fig:studyformatfigure}
    \Description{Illustration of the three studies. For the performance study, there are arrows going from 51 UI mockups to 3 raters to 51 UIs x ~5 suggestions per UI x 3 raters = 765 ratings. For the heuristic evaluation with human experts study, there are arrows going from 6 UI mockups to 12 Design Experts to Guideline Violations in UI Mockups to Interview. For the usage study, there is a cycle starting with 3 UI mock ups going to 12 design experts going to accuracy and helpfulness ratings going to iterative redesign and then back to 3 UI mockups. There is also an arrow going from this cycle to interview.}
\end{figure}
To explore the potential of GPT-4 in automating heuristic evaluation, we carried out three studies (see Figure \ref{fig:studyformatfigure}). In the Performance study, three designers rated the accuracy and helpfulness of GPT-4's generated suggestions for 51 diverse UI mockups to establish performance metrics across a variety of designs. Next, we conducted a heuristic evaluation study with 12 design experts, who each manually identified guideline violations in 6 UIs. Afterwards, they compared their identified violations with those found by GPT-4 in an interview. Finally, in the Iterative Usage study, another group of 12 designers iteratively refined three UIs each with the tool and discussed how the tool might fit into existing workflows in an interview.
%
We obtained UIs from the Figma Community, where designers share their mockups publicly. To attain a diverse set of UIs, we searched for UIs from various app categories, such as finance and e-commerce. We selected UIs that have room for improvement (based on our guidelines) and have JSON representations that could fit into GPT-4's context window. We only used mobile UIs because web UIs were usually too large. For each UI, we ensured that the grouping structure in the Layers panel matched the visual grouping structure in the UI screenshot. We also used our tool to automatically generate semantically informative names for unnamed groups in the Layers panel. 

\subsection{Performance Study}
We recruited three designers for the Performance study through advertising at an academic institution. Each participant had 3-4 years of design experience, and their areas of expertise include mobile, web, product, and UX design. This background information was collected during a brief instructional meeting conducted prior to participants starting this task. We precomputed the guideline violations for all 51 UIs to ensure that all participants saw the same suggestions, allowing us to calculate inter-rater agreement.
The 51 UIs were split into three groups of 17, and each group was evaluated using one set of guidelines. 
Each participant saw the same set of 51 UIs and were given a week to rate the suggestions. Participants spent an average of 6.8 hours total on this task. 

For each suggestion, participants were asked to select a rating for accuracy on a scale of 1 to 3 (``1 - not accurate'', ``2 - partially accurate'', ``3 - accurate'') and then provide a brief, one-sentence explanation for their rating.
Participants were also asked to rate the suggestion's helpfulness on a scale of 1 to 5, with 1 being ``not at all helpful'' and 5 being ``very helpful'', and also provide a brief explanation. We stored all GPT-4 suggestions, along with the corresponding anonymized rating data, explanations, and UI JSONs from this study, and have made this dataset available in the Supplementary Materials.

\subsection{Manual Heuristic Evaluation Study with Human Experts}
We recruited 12 participants through advertising at a large technology company and an academic institution. Two participants had less than 3 years of design experience, six had 3-5 years, two had 6-10 years, and one had 15 years. Their areas of expertise include mobile, web, product, UX, cross device, and UX and UI research. The study was conducted remotely during a 90-minute session, where participants looked for guideline violations in 6 UIs in a Figma file. Each UI was assigned one of the sets of guidelines for evaluation. 

The first 75 minutes consisted of the heuristic evaluation. Participants were instructed to provide the name of the guideline violated, an explanation of the violation following \cite{sadler1989formative}, and a usability severity rating for each violation found. There were a total of 12 UIs used for this study, and each UI was evaluated by 6 participants. The remaining 15 minutes were allocated for a semi-structured interview, where we demoed the plugin and generated feedback for the same 6 UIs the participant evaluated. We then asked the participants to compare the LLM's violations with their own.

\subsection{Iterative Usage Study}
We recruited another group of 12 participants through advertising at an academic institution and a large technology company. One participant had less than 3 years of design experience, five had 3-5 years, three had 6-10 years, two had 11-15 years, and one had over 32 years. Their areas of expertise include mobile, web, product, UX, mixed reality design, and UX and HCI research. The study was conducted either in-person or remotely during a 90-minute session.
%
Participants were given three UIs in a Figma file, each with their corresponding heuristics assigned for the evaluation.
Participants worked through one UI at a time. They first rated the accuracy and helpfulness of GPT-4's suggestions, following the scales used in the Performance Study. 
However, participants in the Usage study 
were asked to follow helpful suggestions to edit the mockup, though they could skip revisions that require too much work, like restructuring the entire layout. After participants finished revising the UI, they would rerun the plugin to generate a new set of suggestions for the revised mockup and then re-rate the new suggestions. For UIs 1 and 3, participants did one round of edits and two rounds of ratings. For UI 2, participants did two rounds of edits and three rounds of ratings, which is meant to assess the LLM's iterative performance.  
This study used the same set of 12 UIs as the manual heuristic evaluation study (with the same guideline assignments for each UI's evaluation), and each UI was seen by three participants. To assess rater agreement, we again precomputed the first round suggestions for each UI. 
After participants finished all three tasks, we concluded with a semi-structured interview, focusing on overall impressions, potential drawbacks and dangers, potential for iterative use, and fit with their design workflow.

\section{Results}
\subsection{Quantitative Results: Performance Study}
\begin{figure*}[t]
 \includegraphics[width=\textwidth]{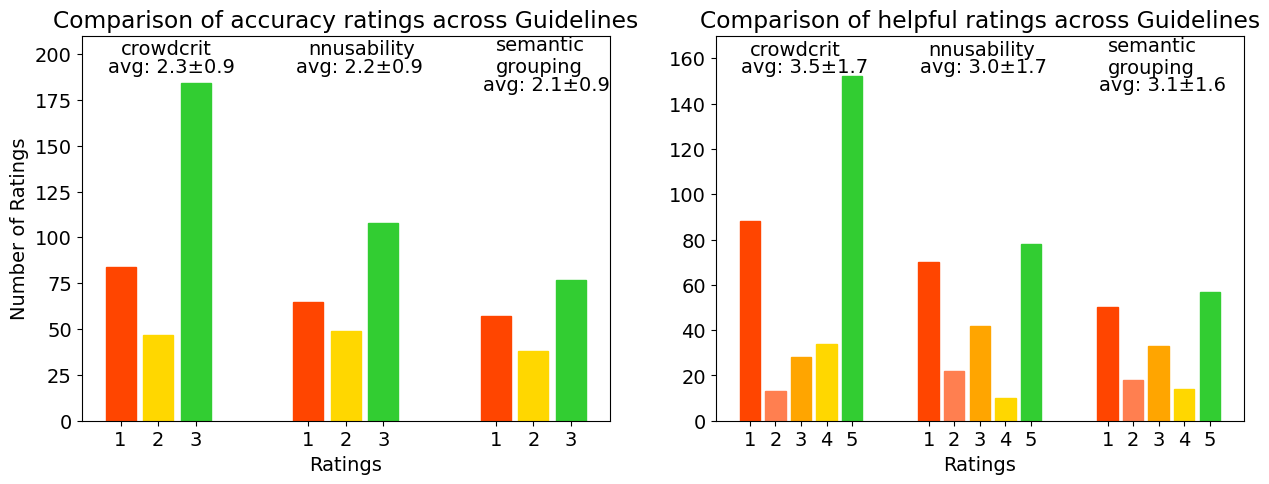}
  \caption{Histogram showing the number of ratings in each category for accuracy and helpfulness, from the 3 participants in the Performance Study. For accuracy, the scale is: ``1 - not accurate'', ``2 - partially accurate'', and ``3 - accurate''. The scale for helpfulness ranges from ``1 - not at all helpful'' to ``5 - very helpful''. The rating data is also visualized as horizontal bar charts for this study and the Usage Study.}
  \label{fig:performancehistogram}
\Description{Two histograms, one showing the accuracy ratings for each category, broken down by guidelines, and the other showing accuracy ratings. The majority of the ratings are concentrated at the two extremes (e.g. accurate and not accurate).}
\end{figure*}
\begin{figure*}[t]
  \includegraphics[width=\textwidth]{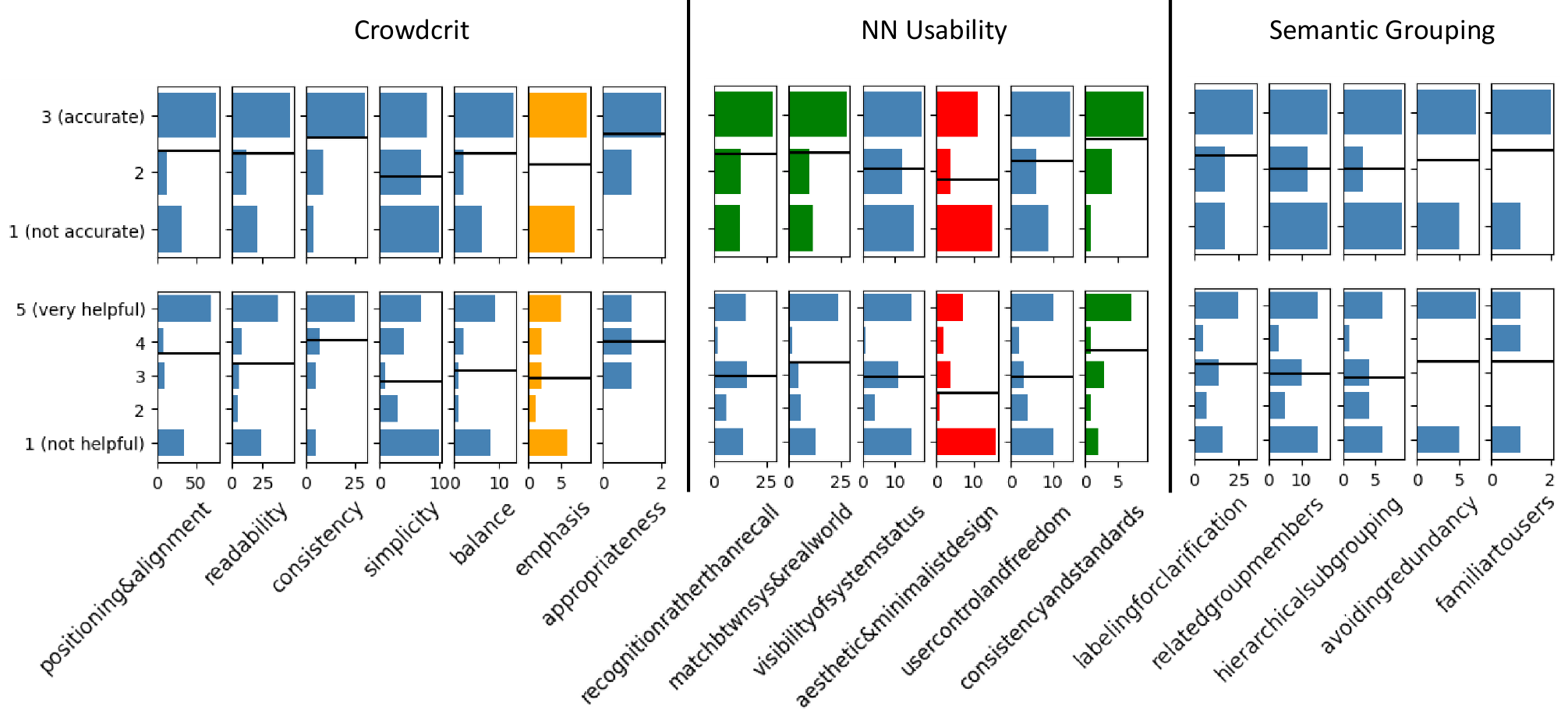}
  \caption{Horizontal bar charts showing the distribution of ratings from the Performance Study for each individual guideline. The ratings for accuracy are in the top row, and helpfulness is in the bottom row, and each chart has a horizontal black line depicting the average rating. We highlight several guidelines with high ratings in green, such as ``Consistency and Standards'' from Nielsen Norman's 10 Usability Heuristics. We used orange to highlight an average performing guideline -- ``Emphasis'' (from CrowdCrit), which had bimodal ratings for accuracy and helpfulness. Finally, we used red to highlight the worst performing guideline -- ``Aesthetic and Minimalist Design'', which had generally poor accuracy and helpfulness ratings.}
  \label{fig:guidelineviolinperformance}
  \Description{Three horizontal bar plots showing the distribution of accuracy and helpfulness ratings for each set of guidelines broken down by individual guideline. The accuracy ratings for recognition rather than recall, match between system and real world, and consistency and standards are highlighted in green because they are generally accurate. The consistency and standards helpfulness rating is also highlighted in green because it is generally helpful. The emphasis guideline is highlighted in yellow for both accuracy and helpfulness because it is bimodal, and aesthetic and minimalist design is highlighted in red for both ratings because it is generally negative.}
\end{figure*}
More GPT-4 generated suggestions were rated as accurate and helpful than not. Across all generated suggestions, 52 percent were rated Accurate, 19 percent Partially Accurate, and 29 percent Not Accurate; 49 percent were considered helpful or very helpful, 15 percent moderately helpful, and 36 percent slightly or not at all helpful. We show histograms of ratings given to all suggestions for each set of guidelines in Figure \ref{fig:performancehistogram}, along with averages.
In line with the aggregate statistics, GPT-4 is more accurate and helpful than not for each set of guidelines.
Furthermore, this difference is largest for CrowdCrit's visual guidelines and smallest for the Semantic Grouping guidelines. Regarding the average rating for all suggestions, CrowdCrit outperformed the other guidelines for accuracy and helpfulness, with a greater outperformance in helpfulness. Semantic Grouping had the worst performance for accuracy, and Nielsen Norman performed the worst for helpfulness. Later, we show concrete examples of GPT-4 generated feedback in Figure \ref{fig:goodandbadsuggestions} that includes accurate and helpful suggestions, as well as inaccurate and unhelpful ones.

We also grouped the ratings by individual guideline and visualized them in horizontal bar charts shown in Figure \ref{fig:guidelineviolinperformance}.
This reveals finer-grained types of heuristics on which GPT-4 performed better than others. The accuracy was highest for the ``Recognition rather than Recall'', ``Match Between System and Real World'', and ``Consistency and Standards'' usability heuristics (highlighted in green in Figure \ref{fig:guidelineviolinperformance}), and participants generally found ``Consistency and Standards'' violations helpful. ``Consistency and Standards'' mostly caught inconsistencies in the visual layout of the UI, like misalignment and inconsistency in size. In contrast, the ``Aesthetic and Minimalist Design'' guideline was generally inaccurate and hence unhelpful, as shown in red in Figure \ref{fig:guidelineviolinperformance}. Finally, the ``Emphasis'' principle (shown in orange) from CrowdCrit was bimodal -- a fairly even distribution of ``accurate'' and ``inaccurate'', as well as ``very helpful'' and ``unhelpful ratings''. The ``Emphasis'' principle mostly identified issues related to the visual hierarchy of the UI.

The subjective nature of heuristic evaluation was already highlighted by Nielsen~\cite{Nielsen1990HeuristicEO}. To characterize subjectivity in our study, we computed inter-rater reliability using Fleiss' Kappa \cite{fleiss1971mns}. Accuracy ratings had an agreement score of 0.112 and helpfulness ratings had a score of 0.100, which suggests only slight agreement. 
In addition to the subjective nature of this task, particular choices in the phrasing of the suggestions could have also lowered agreement scores. For example, the suggestion ``The icons in this group ... could be more user-friendly with the addition of text labels'' calls for a subjective opinion on whether or not text labels are needed, and raters might reasonably disagree.

\begin{table}[t]
\centering
\begin{tabular}{ccc}
\toprule
\textbf{Performance Metrics} & \textbf{GPT-4} & \textbf{Human Evaluator (Avg.)}\\ \midrule
Precision & 0.603 & 0.829\\ \hline
Recall & 0.380 & 0.336\\ \hline
F1 & 0.466 & 0.478\\ 
\bottomrule
\end{tabular}
\caption{Table showing the Precision, Recall, and F1 scores of GPT-4 and an individual human evaluator, computed from the ground truth dataset. The metrics for the human evaluator is computed by averaging these metrics across all participants in the study (for the 6 UIs they each evaluated).}
\label{tab:llmvshumanperformancemetrics}
\end{table}

\subsection{Quantitative Results: Comparison with Human Evaluators}\label{groundtruth}
We compiled all violations identified by the 12 experts from the manual heuristic evaluation study. In total, the experts found 72 distinct guideline violations in the 12 UIs. However, participants sometimes combined multiple violations for a single group or element (e.g., ``The spacing between each icon is inconsistent, and the entire bottom menu is not centered.''). After splitting these combined violations into separate issues, the count increased to 91 distinct violations. GPT-4 found 38 helpful violations for the same set of 12 UIs (determined from the Usage study). 
Nine of these violations were missed by human experts, so in total, the experts and GPT-4 found 100 distinct violations.
In summary, 9 violations were found by GPT-4 only, 29 were found by both GPT-4 and human experts, and 62 were found by human experts only.

We built a ground truth dataset consisting of these 100 violations and computed precision, recall, and F1 performance metrics for GPT-4, which can be found in Table \ref{tab:llmvshumanperformancemetrics}. We also compute the average performance for an individual human evaluator, by averaging these metrics across all participants, which can also be found in Table \ref{tab:llmvshumanperformancemetrics}. On average, a human evaluator had higher precision than GPT-4, which means the guideline violations they found were more likely to be helpful. The average human precision is less than 1, because study participants sometimes found issues irrelevant to the set of heuristics used (e.g., recorded visual grouping issues when using the Semantic Grouping guidelines). GPT-4 scored slightly higher in recall and 
slightly lower in the F1 score than the average human evaluator, though both these differences are much smaller than the difference in precision.

\subsection{Quantitative Results: Iterative Usage Study}
\begin{figure*}[t]
\includegraphics[width=\textwidth]{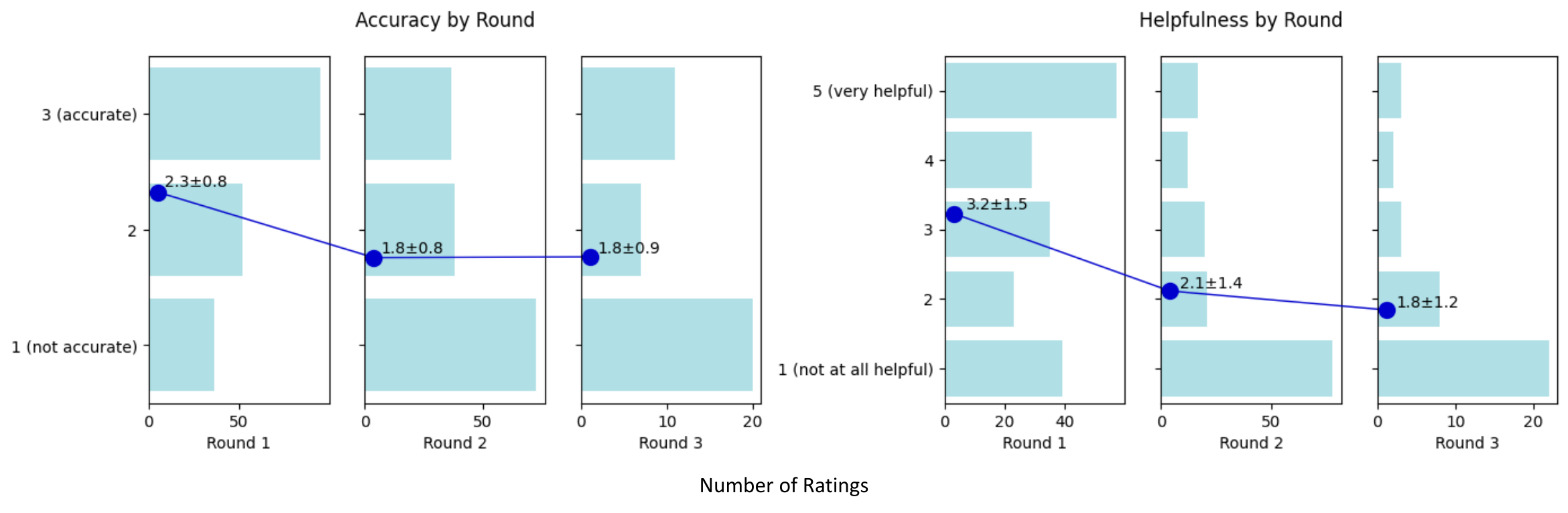}
  \caption{Horizontal bar charts showing the distribution of ratings for each round of evaluation in the Usage study. The ratings are for suggestions from all participants and sets of guidelines. The average rating and standard deviation is marked for each round, and there is a general downward trend in performance as the number of rounds increases.}
  \label{fig:avgratingperround}
  \Description{Horizontal bar charts showing accuracy and helpfulness ratings across rounds. There is a general decrease in performance as the rounds increase.}
\end{figure*}
We collected accuracy and helpfulness scores in the Usage study. Statistics on ratings for suggestions on initial UIs, before participants made any changes, closely match that of the Performance Study: 52 percent were rated Accurate, 28 percent Partially Accurate, and 20 percent Not Accurate; 47 percent were considered helpful or very helpful, 19 percent moderately helpful, and 34 percent slightly or not at all helpful (see Figure~\ref{fig:avgratingperround}).

However, score distributions were lower for later rounds, after participants edited the UIs.
Namely, 39 percent were rated Accurate, 26 percent Partially Accurate, and 35 percent Not Accurate; 33 percent were considered helpful or very helpful, 16 percent moderately helpful, and 51 percent slightly or not at all helpful.
This discrepancy suggests that participants' opinions of suggestions changed during the iterative re-design process. To investigate this, we examined the ratings given per round of iteration in the Usage Study.
Since GPT-4's suggestions vary in accuracy and helpfulness, we used a horizontal bar chart to show the distributions of all ratings given to suggestions in each round. Figure \ref{fig:avgratingperround} shows a general trend of decreasing performance per round for both accuracy and helpfulness. This trend also generally holds for both metrics when broken down by guidelines and participants (available in the Supplementary Materials).

The average inter-rater reliability score, based on the first round of suggestions, is 0.155 for accuracy and 0.085 for helpfulness, which again indicates subjectivity in the experts' opinions towards the suggestions.
\subsection{Qualitative Results: GPT-4 Strengths and Weakness}\label{llmweakness}
\begin{figure*}
  \includegraphics[width=\textwidth]{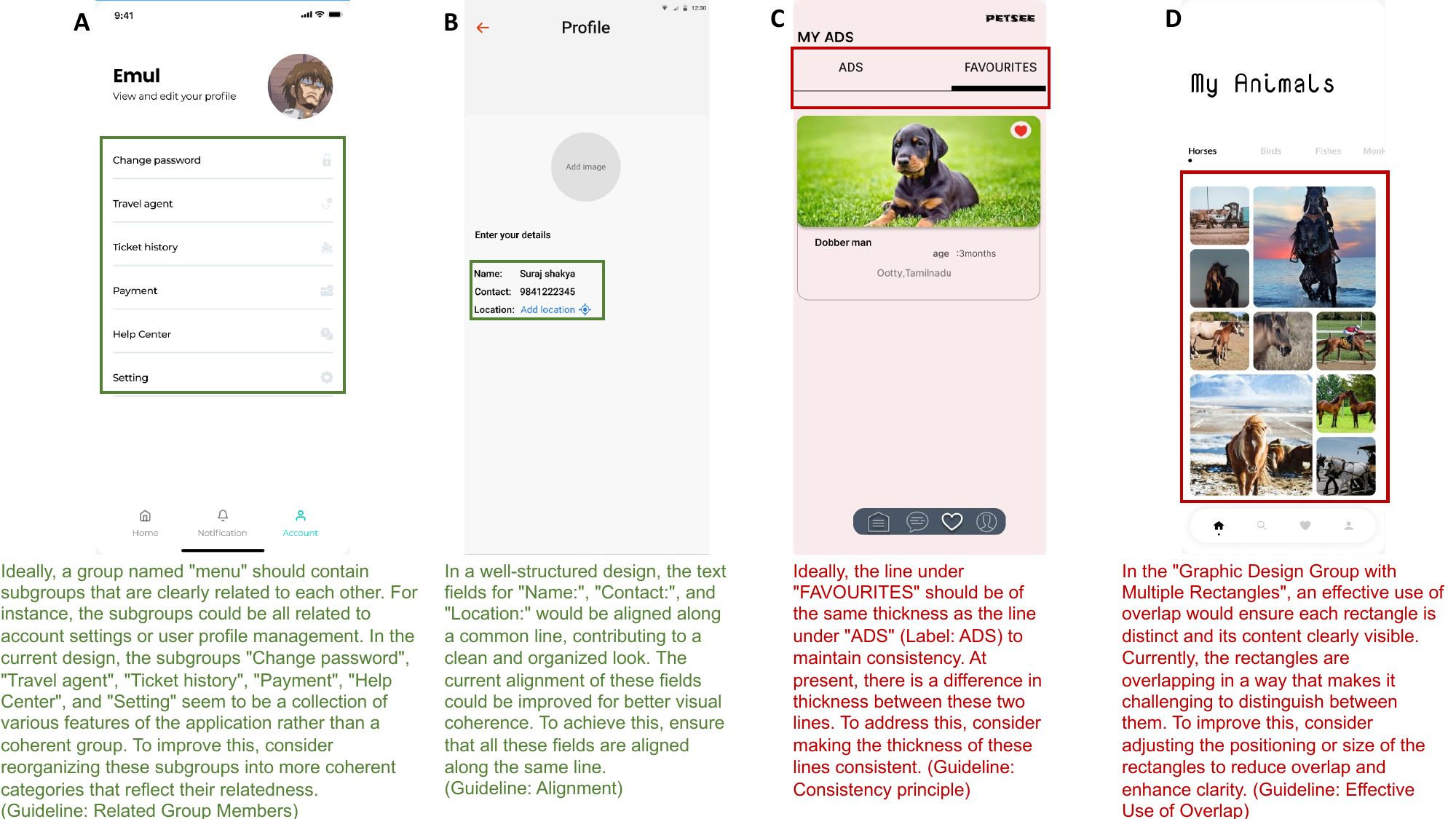}
  \caption{Examples of GPT-4 suggestions that all participants found very helpful or unhelpful, along with their corresponding UIs above (with the relevant group marked). The suggestions for UIs A and B received ratings of 5 for helpfulness and were rated as accurate by all three participants (from the Usage study). The ``Contact:'' field for UI B is slightly misaligned from the other fields, which GPT-4 caught. UIs C and D were rated 1 for helpfulness by all three participants. For UI C, the LLM stated that the line thickness was uneven under the ``ADS'' and ``FAVORITES'' tab, which is technically accurate (and some participants rated it as accurate) but unhelpful as the uneven line thickness is meant to indicate the selected tab.}
  \label{fig:goodandbadsuggestions}
  \Description{Four examples of GPT-4 suggestions with corresponding annotated UI screenshots. The two on the left are highly rated with and discuss further subgrouping a menu with unrelated items and aligning a slightly misaligned form. The two on the right are low rated. One says that the line thickness is uneven for a set of tabs what show which page is displayed, and the other says that a grid of images are overlapping when they are not visually.}
\end{figure*}
We analyzed GPT-4's suggestions, corresponding expert ratings, explanations, and interview responses from the Usage study. Through grounded theory coding \cite{GlaStr67} of the qualitative data and subsequent thematic analysis \cite{thematicanalysis}, we identified the following emerging themes on GPT-4's strengths and weaknesses. Figure \ref{fig:goodandbadsuggestions} contains examples of high and low-rated LLM suggestions to illustrate some of these themes.


\subsubsection{Strength 1: Identification of Subtle Issues (12/12 Participants)}
All participants found GPT-4's ability to identify subtle, easy-to-miss issues helpful. This includes problems like misalignment, uneven spacing, poor color contrast, redundant elements, and uncommon icons without text labels. UI B in Figure \ref{fig:goodandbadsuggestions} contains an example of a misalignment caught by GPT-4 that all three participants found very helpful. Seven participants mentioned this theme as a strength during the interview. For instance, P5 stated that the tool is ``useful for pointing out things that are not obvious to the naked eye, like minor visual details'', and P4 said that they ``liked the UI at first, but then the LLM found small issues with labels, etc.''

\subsubsection{Strength 2: Fixing Text-related Issues in the UI (12/12 Participants)}
GPT-4 was also effective in identifying text-related issues, like incorrect grammar, unclear text labels, and text that is not user-friendly (e.g., uses too much jargon). In addition, GPT-4 would usually include the correct text to use in its feedback. For instance, one UI had a grammatically incorrect header: ``what kind of pet your Looking?'' and GPT-4 suggested revising it to ``What kind of pet are you looking for?'' P1 rated this suggestion as very helpful and explained that ``it gave me the right content to copy and use in the design.''

\subsubsection{Strength 3: Reasoning with UI Semantics (8/12 Participants)} 
GPT-4 was skilled at reasoning with UI semantics. This includes identifying large groups of items that could be subgrouped into smaller, more semantically-related groups. This is illustrated in UI A of Figure
\ref{fig:goodandbadsuggestions}, where the LLM suggested subgrouping the menu items into more related categories. P2 commented ``Yes, I think it would be helpful to separate out the sections into categories, as some relate to the purpose of the application (traveling), and other parts are related to basic maintenance of the account (e.g., settings)''. Other semantics-related issues that GPT-4 identified include finding groups or subgroups with elements are not clearly related, groups or visualizations where the purpose is unclear and requires a text label to explain them, and confusing UIs that require documentation.  

\subsubsection{Other Strengths}\label{llmotherstrengths}
For two participants, GPT-4 made clever, high-level suggestions. For instance, GPT-4 recommended that a product quantity value be changed to an editable field, so users will not have to rely on the `+' or `-'  buttons to adjust the quantity, and cited Nielsen's ``Flexibility and Efficiency of Use'' heuristic. In addition, GPT-4 suggested that a list of reviews for a cafe should display aggregate statistics, like the total number of reviews or an average star rating. P1 encountered both these suggestions and stated that GPT-4 ``found two good ones that I would think about -- with reviews and product quantity input field''. P12 encountered a violation where the selected tab in the navbar was called ``strategies'', but the displayed page was about stock performance, which was unrelated to strategies. They commented that GPT-4 was ``spot on''.

\subsubsection{Weakness 1: Overapplication of Guidelines (12/12 Participants)}
GPT-4 sometimes over-applied guidelines too literally, without considering the context provided by the rest of the UI, popular design conventions, nor conflicts with other guidelines. UI C in Figure \ref{fig:goodandbadsuggestions} shows an example where the LLM correctly identified that the line thickness is inconsistent for the two tabs; however, this was a conscious design decision to show the selected tab in the broader context of the UI. P7 commented ``the line under favorites indicates that we are on the favorites tab/screen; making it consistent would eliminate this distinction''. Regarding popular design conventions, GPT-4 often recommended labels for common icons, like the `X' and shopping bag icons, but ``universal icons do not need to be labeled'', as stated by P1.

\subsubsection{Weakness 2: Repetition of Feedback (6/12 Participants)}
GPT-4 sometimes repeated the same feedback for every element of the same type. For instance, in the second UI of Figure \ref{fig:goodandbadsuggestions}, the LLM suggested increasing the spacing between the ``Name:'' label and the name ``Suraj shakaya'' and then repeated this same suggestion for both ``Contact:'' and ``Location:''. P4 commented that the LLM ``repeats the same type of issue, and it is especially annoying when it is incorrect.''

\subsubsection{Weakness 3: Limitations of the JSON Representation (8/12 Participants)}
A limitation of using JSON to represent the UI is that GPT-4 cannot capture violations that would require processing the rendered image of a UI. For example, GPT-4 will call out issues with elements overlapping when there is no visual overlap.
UI D in Figure \ref{fig:goodandbadsuggestions} is an example where the LLM flagged an overlap issue because the bounding boxes overlap, but the photos do not overlap visually. 
Another error is that the LLM does not recognize center alignment for items of different sizes. 
This is explained by P9: ``The elements are of different sizes and the bounding boxes of elements are not aligned, but the inner contents of the boxes are visually aligned.''

\subsubsection{Weakness 4: Vague Suggestions (5/12 Participants)}
Five participants stated that GPT-4's suggestions were too vague, specifically regarding how to fixing the violation. P1 stated that the LLM would ``be more useful if it were more specific and gave more examples on how to fix things'', and P11 said they would ``like to see more actionable suggestions, like change this color to a specific value''. The reason for this vagueness is due to our system implementation. We first attain a set of guideline violations from GPT-4 and then send the violation explanations only (without the UI) for the LLM to rephrase into constructive feedback. Since GPT-4 does not have information on the UI, it cannot suggest specific fixes in the UI to address each violation. We tried passing in the UI for the rephrasing call, but it led to high latency (several minutes), so we decided to not include the UI JSON from a usability perspective.

\subsubsection{Other Weaknesses}
We noticed an interesting behavior of GPT-4 in response to designer feedback. When the LLM is notified that the guideline violation is incorrect for a group or element (from the designer hiding the suggestion), it would sometimes select a different guideline to explain the same issue. For instance, GPT-4 would often cite a violation of ``Recognition over Recall'' for unlabeled icons, but if the designer dismisses this suggestion, GPT-4 would repeat the same suggestion in the next round and cite ``Match Between System and Real World'' instead. 
This behavior was encountered by 5 participants. 

\subsection{Qualitative Results: Comparison with Human Evaluators}
\begin{figure*}
  \centering
  \includegraphics[width=\textwidth]{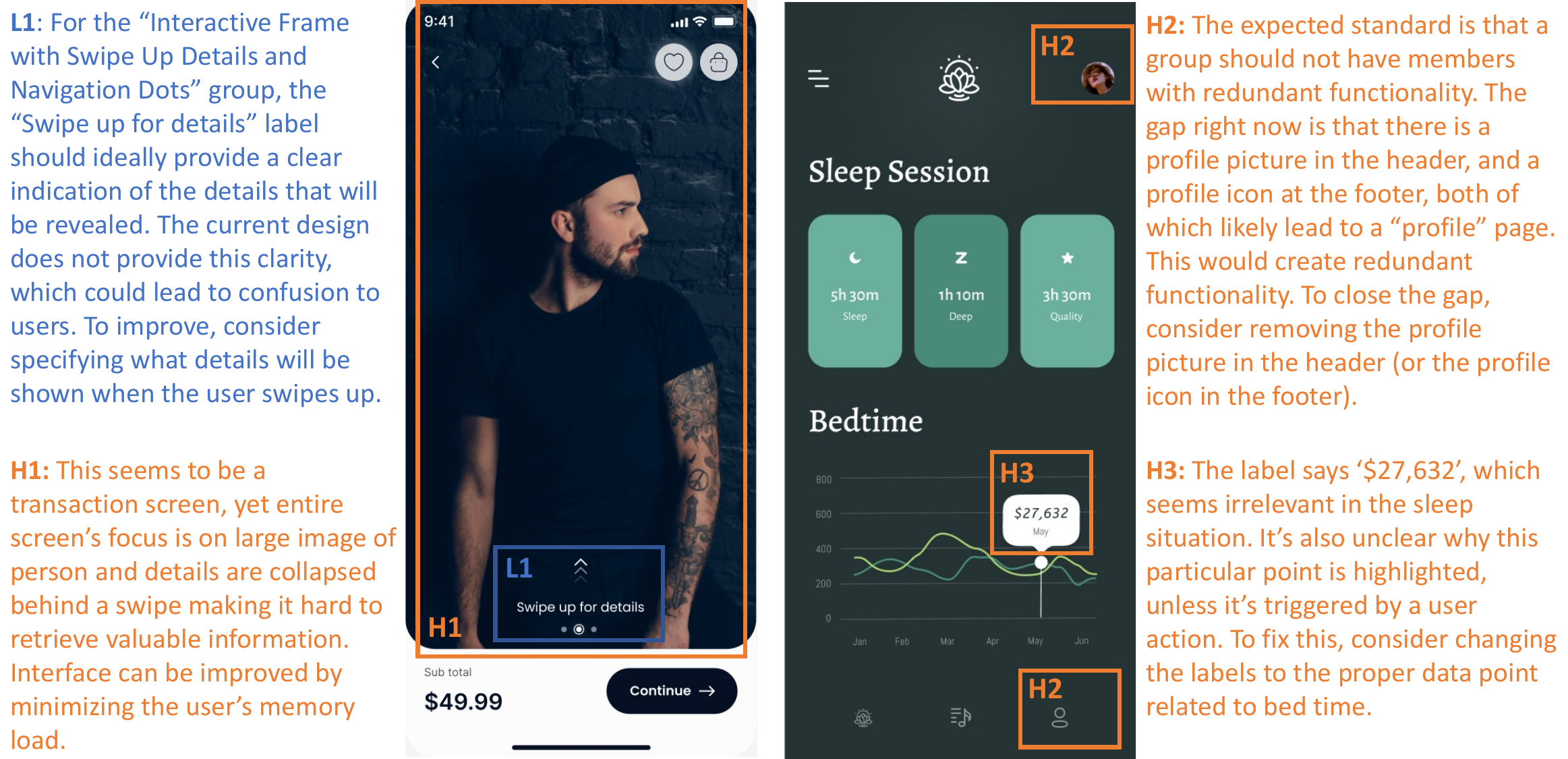}
  \caption{The screenshot on the left compares a high-level ``global'' violation found by a human expert with a similar, but more specific, violation found by the LLM. The right screenshot contains two violations found by human experts that require advanced visual understanding of the UI.}
    \label{fig:comparisonwithhuman}
  \Description{Two screenshots with violations marked. The left screenshot has a box draw around the large image taking up most of the UI and the corresponding violation for this box describes how this large image hides important details of the UI. Another box is draw around the "Swipe up for details" text label and the corresponding violation talks about how the label should be edited to indicate what kind of details will be revealed. The right screenshot has a box drawn around a profile picture and a profile icon and it maps to a violation regarding redundant links to the profile, and the right screenshot also has a box draw around a tooltip in the graph, which talks about how the tooltip does not match the content nor axes of the graph.}
\end{figure*}

We used grounded theory coding and thematic analysis to characterize key qualities for 1) issues violations found by GPT-4 only, 2) found by both GPT-4 and human experts, and 3) found by human experts only. 
Finally, we summarize participants' feedback for the LLM from the concluding interview.

\subsubsection{Violations Found by GPT-4 only (9 percent)}\label{llmonly}
A small fraction of the violations were found by the LLM only. Eight of the nine issues were quite subtle, covering poor text contrast, text labels that require clarification, and misalignment. The ninth suggestion missed by all participants recommended adding a new feature to filter through the cafe's reviews, and participant H8 commented ``this type of suggestion did not cross my mind during the manual evaluation''.

\subsubsection{Violations Found by both Humans and GPT-4 (29 percent)}
Twenty-nine issues were identified by both humans and the LLM. The majority of these violations (15 violations) involved text labels -- labels that require clarification, and missing labels to explain an icon, element, or group. The second most common issue (6 issues) involved positioning and alignment, with four being quite obvious (e.g., the misaligned elements were scattered like in Figure \ref{fig:goodandbadsuggestions}C). Other issues found by both include problems with the UI text (e.g., the text had too much jargon) and the hierarchical subgrouping violation shown in Figure \ref{fig:goodandbadsuggestions}A.

\subsubsection{Violations Found by Humans only (62 percent)\label{violationshumansonly}}
The majority of the violations were found by human experts only (62 violations). Common characteristics of violations in this category include ``global'' (i.e., high-level) issues of the UI, advanced visual issues, and violations consisting of multiple distinct issues. Participants found 11 high-level issues that require understanding of the UI's purpose and context. 
Figure \ref{fig:comparisonwithhuman} (Violation H1) illustrates a global violation identified by a human expert, where a large image hides important content, and compares it to a similar, but more specific, violation found by GPT-4 (L1). While GPT-4 is able to find these ``global'' violations, as discussed in Sections \ref{llmonly} and \ref{llmotherstrengths}, human experts found considerably more. Eight violations, found only by human experts, required advanced visual understanding of the UI. The right screenshot in Figure \ref{fig:comparisonwithhuman} illustrates two such violations. The tooltip (Figure \ref{fig:comparisonwithhuman} H3) displayed a monetary amount that not only exceeded the axes of the graph but also mismatched the graph's content about sleep duration. Violation H2 highlighted redundant links to the user profile with via both a profile image and a profile icon. Finally, a few participants stated that parts of the UI shown in Figure \ref{fig:crowdcritexampleuiovertime} (Round 1) had clashing visual design and an overly complicated background.

Ten of the violations recorded by the participants involved combinations of several distinct issues for a single group or element. For instance, a violation for the UI in Figure \ref{fig:crowdcritexampleuiovertime} (Round 1) stated that ``The title has incorrect spelling and grammar, is not aligned on the page/has awkward margins, has inconsistent text styles for the same sentence, and includes clashing visual elements''. 
Finally, participants found 22 issues that were similar to the types of issues caught by GPT-4, such as misalignment, unclear labels, and redundancy. This implies that GPT-4 is less comprehensive than a group of 6 human experts, as each UI was evaluated by 6 study participants.

\subsubsection{Interview Findings}
Participants were generally impressed by the convenience of this plugin, which could find helpful guideline violations at a much faster speed than manual evaluation. H6 said that it ``can cut about 50 percent of your work, and is at the level of a good junior designer'', and H3 said they ``wish it was already out for use''. Compared to the violations they found manually, several participants said GPT-4 was more thorough and detailed (H3, H4, H5, H6, H10). H1 ``appreciated how the LLM could find subtle violations that were missed'', and P5 said they were ``overwhelmed by the number of issues in some UIs'' and appreciated how the LLM can catch violations that were ``tedious to find''. H6 said GPT-4 ``goes into a much lower level of resolution than is commercially feasible to do, since it takes a long time''. H1, H3, and H9 valued how GPT-4 could sometimes better articulate the violation. H9 stated that they were ``pleasantly surprised at how it picked the right way to describe the problem'', regarding an issue they struggled with describing. Finally, H1, H2, H4, H7, and H8 all appreciated how GPT-4 found violations that were missed during their manual evaluation. H7 said ``it was useful, as it captured more cases than I found''.

Participants brought up weaknesses of GPT-4's feedback, which mostly aligned with the findings in Sections \ref{llmweakness} and \ref{violationshumansonly}. These limitations include missing the majority of the ``global'' violations (H1, H5, H9), limited visual understanding of the UI (H2, H8, H11), and poor knowledge of popular design conventions (H2, H7, H8, H10, H11). Finally, like the participants in the Usage study, those in this study also did not consider the LLM’s mistakes to be a significant issue. H10 said ``if the feedback is correct, then is it very helpful, and if not, it is not a big deal as you can just dismiss it'', and H6 said ``the 60 percent success rate is not a problem, as it saves a lot of time in the end''.

\subsection{Qualitative Results: Integration into Existing Design Practices}
We analyzed the interview responses from the Usage study with grounded theory coding and thematic analysis to determine this tool's fit into existing design practice. The emerging themes centered around how and when designers would integrate it in their practice, potential broader use cases, and possible dangers of an imperfect tool.

\subsubsection{How and When Designers Would Integrate this Tool in Practice}
Nine out of 12 participants said they would use this tool. The three who would not cited inaccurate and vague suggestions as reasons. Of the nine participants who would use this tool, three said they would use it during the initial stages of the design. This includes tasks like determining the hierarchy of components in low-fidelity prototypes (P5, P7) and ``exploring early stage concepts'' (P7).
%
The other six participants would use this tool in the later design stages. P5 said they would also use it ``after the first draft of high fidelity for alignment issues''. P11 said ``after finalizing their design, I would run a check before sending it to engineering'', and they would also run this tool ``after making large design changes to see if any design considerations were missed''. Finally, P6 said they would run this tool ``after prototyping and before usability testing; it is valuable to test the LLM's recommendations during the user test''.
%

Regarding iterative or one-shot usage, seven participants preferred using this tool for a single check, and the other five preferred iterative usage. The participants who preferred one-shot usage stated that the LLM feedback was less accurate and helpful in later rounds (P4, P5, P7), found that the ``first round of edits was sufficient'' (P8), or preferred to use the LLM's suggestions as ``tips to start off the thinking process, but not rely on it'' (P7).

Participants also had positive feedback on the interaction design of the plugin. Six participants stated that they liked being able to click on the group or element's name in the suggestion to select it in the mockup (Figure \ref{fig:plugininteractions}, A).
Four participants stated they like the references to guidelines, as it ``adds credibility'' (P3), and they could reference the guideline's source material for more context (P9). P3 liked the constructive framing of the violations, saying that ``the type of language is encouraging, which is nice to see''. Finally, P4 liked the ``built-in guideline options and the flexibility to specify my own guidelines''. 

\subsubsection{Potential Broader Use Cases}
The participants brainstormed a long list of use cases for this tool. Some use cases are based on real observations from the Usage study, and others are hypothetical (from speculation). For observed use cases, the most common was using this tool to save busywork, which was mentioned by 5 participants. P10 said this tool would ``save time on mundane tasks'', for things like ``spellcheck'', and 3 participants said it can catch small details like line spacing, size, and ``small things you might miss'' (P11). Other observed use cases include evaluating other people's designs during teamwork (P3) and secondary research (P8), and using it ``as a first round of usability testing'' (P8). 
Common hypothetical use cases include using it for accessibility checks (4 participants) and training novices (10 participants). P5 stated that this tool can ``help younger designers learn and reinforce these guidelines'', and P11 and P12 said the mistakes made by the LLM can train novices to carefully consider design suggestions and be more skeptical. Other speculative use cases include getting a second opinion on your UIs (for designers working solo) (P6), checking for compliance with brand standards (P10) and company rules (P12), and ``large-scale evaluation, where you designed a lot of screens and want a quick evaluation'' (P9).

\subsubsection{Potential Dangers of this Tool}
Although the tool sometimes reported inaccurate violations, most participants (8 out of 12) considered the tool not dangerous because there is a human in the loop to catch errors. However, some potential dangers stated by participants include ``thinking there is something wrong with the design when there might not be'' (P2) and users who trust the LLM 100 percent (P1, P4, P6, P9). For instance, novices may fully follow the LLM's advice (P1, P2, P3, P6, P11), and P3 suggested that they could use this tool with expert supervision. However, 4 participants felt that novices should be able to detect the LLM's mistakes. P3 stated that ``the wrong suggestions are outlandish enough that novices can tell that it does not make sense''. There were also some negative feedback regarding the plugin design. Three participants (P2, P8, P11) found the LLM suggestions too wordy, and P7 did not like how the tool ``can only evaluate one screen at a time, as opposed to the whole flow.'' 

\section{Discussion}
\begin{figure*}
  \includegraphics[width=\textwidth]{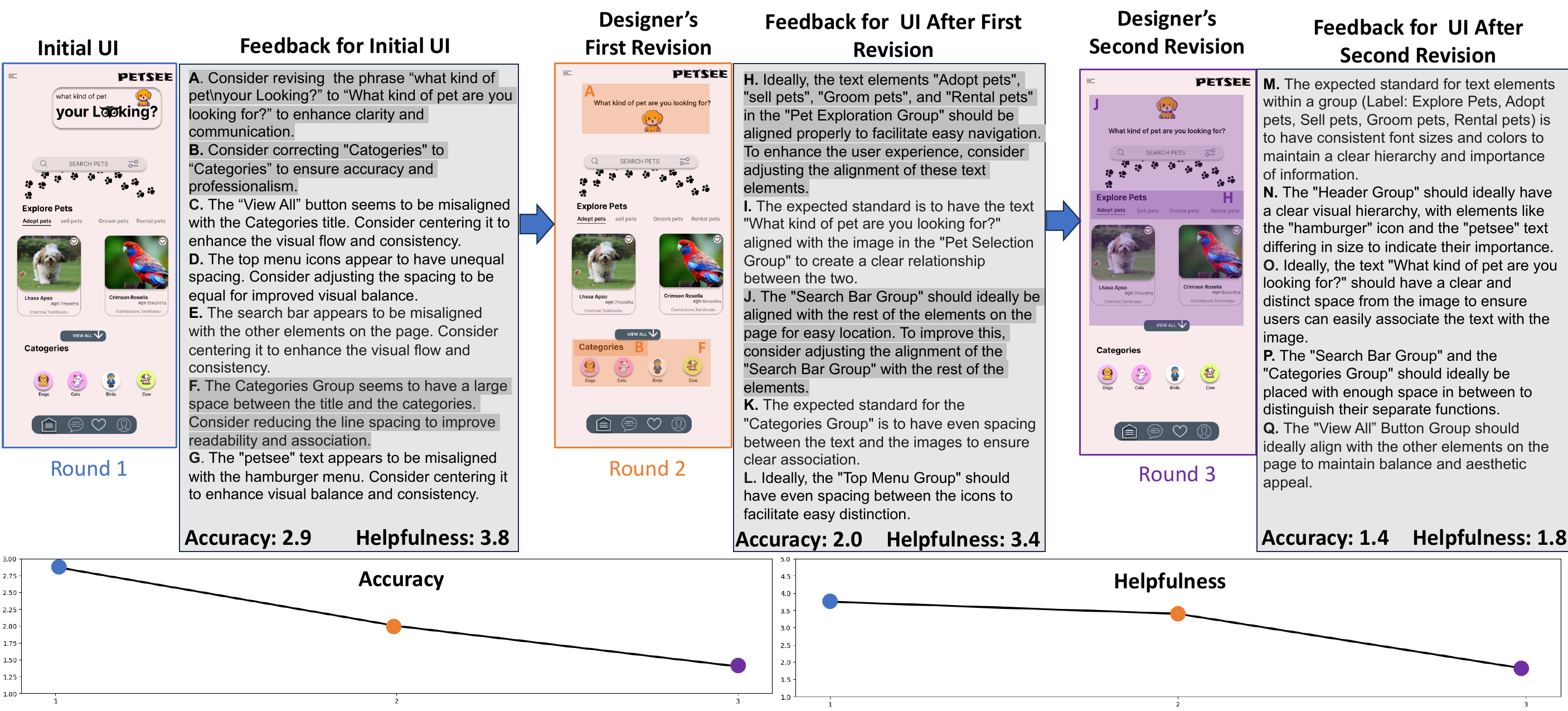}
  \caption{Plots of the average accuracy and helpfulness rating per round from P2, who used the CrowdCrit guidelines to evaluate the UI shown above. Paraphrased GPT-4 suggestions from each round and their average accuracy and helpfulness ratings are also provided. The suggestions P2 used to edit the UI are highlighted in gray, and the corresponding changes are annotated in the revised UI. Participants made at most 2 rounds of edits, so the ``Feedback for UI After Second Revision'' was never used for edits. Note that participants were instructed to approach this task as if they were using this plugin for their own design work and to put as much effort into the edits as they would like. This sometimes led them to fix issues in the design beyond what the LLM explicitly stated. 
  The UI's visual design improves per round, while the average accuracy and helpful ratings decrease.}
  \label{fig:crowdcritexampleuiovertime}
  \Description{Three screenshots of the same UI after each round of revision and corresponding LLM suggestions per round. Line plots of the accuracy and performance ratings per round are also included. The UI improves visually across the three rounds while the performance drops. Also, the suggestions that were used by the participant for the revision are also highlighted.}
\end{figure*}
We explored the potential of using LLMs for automated heuristic evaluation. In particular, we assessed the feasibility of the best performing LLM (GPT-4) and how a tool built on this can fit into existing design practice. We discuss interesting insights from our findings, and implications for its feasibility and integration into design practice. 
\subsection{Feasibility of GPT-4 for Heuristic Evaluation}
We assessed GPT-4's performance quantitatively and qualitatively. From our qualitative analysis, we identified a concrete set of strengths and weaknesses. Most of these weaknesses could potentially be addressed, which we will discuss in Future Work (Section \ref{limitations}). 
%
Quantitatively, the GPT-4 was judged to be more accurate and helpful than not during the Performance study and the first round of the Usage study. The authors selected the UIs for both studies because they identified issues according to the chosen guidelines. This suggests that the LLM can identify some weaknesses in poor UI designs. The decrease in performance after each iteration during the Usage study could be explained by the fact that the participants are design experts, so their edits likely improved the UI overall, leaving fewer guideline violations for GPT-4 to detect. As a result, the task becomes harder for the LLM, causing it to detect erroneous violations. Figure \ref{fig:crowdcritexampleuiovertime} illustrates this; GPT-4's suggestions become less accurate and helpful as P2 iterates on the UI using the CrowdCrit guidelines, while the UI's visual design becomes noticeably better going into each round of evaluation. 
This general decrease in performance as the UI improves implies that this tool is perhaps not yet suitable for iterative usage by expert designers. 
In addition to being affected by the quality of the UI mockup, the performance also varies depending on the guidelines used.

\subsubsection{CrowdCrit}
GPT-4 performed the best with CrowdCrit's visual design heuristics in both accuracy and helpfulness. This is likely due to the prevalence of specific visual design heuristics that rely on mathematical checks for alignment, spacing, and consistency in size -- all checks that are straightforward to compute with layout information found in the UI JSON (except for center alignment and overlap). CrowdCrit also covers UI text accuracy (i.e., grammar and spelling), where LLMs excel in identifying and fixing issues. Regarding helpfulness, many visual design errors are subtle (like slight misalignment), and participants found it very useful when the LLM found them. 

\subsubsection{Nielsen Norman 10 Usability Heuristics}
We expected the Nielsen Norman 10 Usability Heuristics to have the best performance, as they are widespread on the web and would have appeared in GPT-4's training data. However, its worse performance, especially for helpfulness, could be because many of its heuristics apply to interactions or flows (like ``User Control and Freedom'', ``Help and Documentation'', and ``Error Prevention'') that are out of scope for single-screen, static mock-ups. In addition, the ``Aesthetic and Minimalist Design'' heuristic covers visual design but had poor performance. This is probably because this heuristic is quite vague, focusing on only presenting necessary information to the user; this is difficult to assess, especially without the context provided by other screens in the flow. On the other hand, GPT-4 performed well on ``Consistency and Standards'', which mostly checked for consistency in the visual layout, like size and alignment, which are straightforward numerical checks.

\subsubsection{Semantic Grouping}
The lack of training data on Semantic Grouping guidelines, which were recently developed, could explain their lower accuracy. In addition, the quality of group/element names in the UI JSON impacts the assessment of the UI's semantic organization. Since designers must manually add these names, they may not always be  accurate. While we can automate the naming of groups, they are determined from the labels of their members. Many leaf elements, like icons and images, are also missing labels, which may affect the accuracy of the automatically generated labels. While GPT-4 generally performed well for evaluating the relatedness of elements, as mentioned by study participants, the times it did not perform well for semantic grouping were probably for poorly annotated UIs. 


\subsection{General Insights into LLMs and their Future Development}
The LLM comparison analysis revealed that this automated heuristic evaluation task of UIs in JSON form is hard enough to require the largest state-of-the-art model with the strongest reasoning skills (GPT-4). The other state-of-the-art models were smaller and struggled with this task; they either missed most of the violations or indiscriminately applied the guidelines and had difficulty following the detailed instructions. While these LLMs found some helpful violations, their performance falls short for building a system. The prompt exploration study showed that breaking tasks into smaller, simpler ones and providing the maximum amount of guidance yielded the strongest performance for GPT-4, and this insight probably applies to other LLMs. 

Furthermore, LLMs are rapidly advancing. Recent developments include multimodal models that can accept images as input (e.g., GPT-4V), and models with much larger context windows (e.g., GPT-4-Turbo). Multimodal LLMs could accept UI screenshots, which may give the model a better visual understanding of the UI; larger context windows would enable more complex UIs, an extensive list of few shot examples, or chain-of-thought prompting. These developments could address some of the limitations we discussed and may improve LLM performance for this task; however, this would have to be evaluated in future work, as our experiments with earlier multimodal models were not successful. 

\subsection{Comparison with Human Evaluators}
While GPT-4 and a human evaluator had comparable F1 scores, 
human evaluators had higher precision, a better understanding of the UI's context, found more ``global'' (i.e., high-level) violations, and had superior visual understanding of the UI. 
GPT-4, on the other hand, was more thorough, detailed, and specific, catching a greater number of helpful issues than an individual human evaluator. Furthermore, GPT-4 was better at catching detailed errors that were tedious for humans to find and was also better at finding subtle violations. These findings imply that the strengths of humans and GPT-4 are complementary for heuristic evaluation.

\subsection{Fit into Design Practice}
We found that even with its current performance, participants were generally positive towards this tool and would use it in their practice. They stated that the errors made by  GPT-4 were easy to detect and not dangerous, as there is a human in the loop to ensure that updates to the UI mockup are based only on valid feedback. 
Given the LLM's strengths, designers have brought up various use cases for this plugin, such as finding subtle visual design errors in their high-fidelity mockups and planning the grouping structure of their low-fidelity designs. The core capabilities of this plugin allow designers to evaluate their Figma mockup against any set of guidelines, and participants suggested extending it to check for accessibility and compliance with company/brand standards. 
The experts' acceptance of the LLM's imperfect suggestions, combined with numerous suggested use cases, implies that an automated LLM-driven heuristic evaluation tool may soon find a place in design practice, supporting human designers with various aspects of the design process.

\section{Limitations and Future Work}\label{limitations}
There are several limitations with this system and studies. 
For the system, the LLM's accuracy on semantic heuristics depends on the  quality of the names designers manually add to elements. Furthermore, due to context window limitations (8.1k tokens for GPT-4), the plugin could only evaluate one static mobile UI screen at a time. This prevents the plugin from evaluating interactivity, design consistency across screens, and task flows. Larger UIs, like websites and desktop apps, are more complex and would also exceed this context window size. While there are models with larger context windows (Claude 2 and GPT-3.5-16k), they did not produce sufficiently helpful output in our exploration. Other LLMs with even smaller context windows, such as Llama 2 and GPT-3.5-turbo, are not suitable for realistic UIs. Regarding the studies, they captured the quality of the LLM's feedback, and participants used the plugin to revise mostly high-fidelity mockups. These studies did not capture the impact of this plugin's usage on a more realistic design scenario, where designers start with an idea and design a UI mockup based on it. 

These limitations suggest exciting opportunities for future work. To start with, we describe potential ways to address the LLM limitations discussed in Section~\ref{llmweakness}. We have stored the data on LLM suggestions, ratings, rating explanations, and UI JSONs from both studies (available in the Supplementary Materials). Future work can use this data to fine-tune an LLM to improve performance. 
Furthermore, to address repetitive suggestions, P5 suggested grouping them into one long suggestion, which could be accomplished by some engineering effort to identify elements of the same type in the UI JSON. 
Future work should also evaluate the performance of emerging multimodal models (e.g., GPT-4V) with a prompt consisting of both the UI screenshot and JSON representation to see if it improves the LLM's visual understanding of the UI. Models with larger context windows (e.g., GPT-4-turbo) should be evaluated to see if they can enable evaluation of task flows across multiple screens. Such models could also enable use of few-shot or chain-of-thought examples. Once these limitations of the plugin have been addressed, a study could be conducted where participants use this plugin to assist in creating a design from a given prompt. This study would more realistically simulate the tool's usage in practice. 


\section{Conclusion}
 We designed and built a Figma plugin that uses LLMs to automate the heuristic evaluation of Figma mockups with arbitrary text-based design guidelines. After determining the optimal LLM for this task (GPT-4), we investigated its capability to automate heuristic evaluation through a study where three design experts rated the accuracy and helpfulness of GPT-4 generated design suggestions for 51 UIs, and also compared its feedback with those provided by human experts. Finally, we explored how this tool can fit into existing design practice via a study where 12 design experts used this tool to iteratively refine UIs, assessed the generated feedback, and discussed their experiences working with the plugin. We found that GPT-4 generally performs well on poor UI designs, but its utility decreases through iterations of revising the design. Participants generally do not find the LLM's mistakes to be an issue, as they are easily detected by human designers. Most participants would already use this tool as part of their design practice. 

\bibliographystyle{ACM-Reference-Format}
\bibliography{./paper.bib}
\onecolumn
\appendix
\section{Appendix}
\begin{figure*}[h!]
  \centering
  \includegraphics[width=0.8\linewidth]{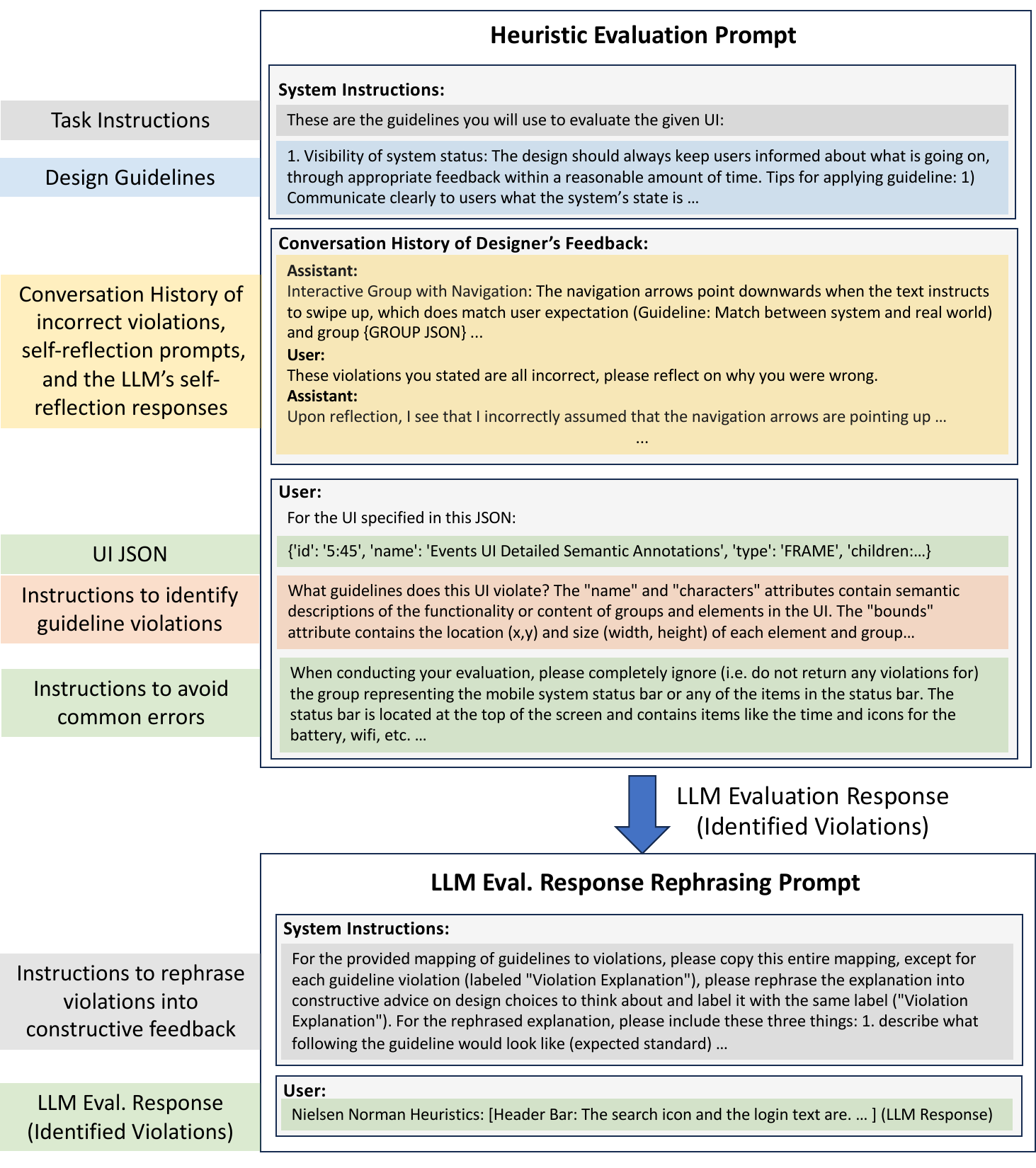}
   \caption{
Diagram illustrating the components of each prompt. The Heuristic Evaluation Prompt and the LLM Eval. Response Rephrasing Prompt form a chain, where the Rephrasing Prompt takes the LLM response from the Heuristic Evaluation prompt and instructs the LLM to rephrase it. The Heuristic Evaluation prompt starts with system instructions to guide the LLM's behavior and contains the set of guidelines for evaluation. It is followed by a conversation history with incorrect/unhelpful violations found by the LLM (as ``Assistant'') that were denoted by the designer, a ``User'' query telling the LLM that these violations were wrong and to self-reflect, and the LLM's response to the self-reflection. There may be zero to multiple sets of this conversation, depending on the number of evaluation rounds. The final component is the user message, which contains the UI JSON, instructions to identify guideline violations, a short description of the content available in the UI JSON, and specific instructions to avoid common errors. The Rephrasing prompt contains system instructions that direct the LLM to constructively rephrase the violation explanation (following \cite{sadler1989formative}) and also guides the LLM to format the response correctly. The user query contains the LLM Eval. response with the identified violations.}
    \label{fig:promptdiagram}
     \Description{Diagram illustrating different components of the prompt. The heuristic evaluation prompt is broken down into system instructions, which contains task instructions and the design guidelines. The Conversation history of the designer’s feedback contains a conversation about incorrect violations found by the assistant and the user telling the assistant to self reflect, along with the assistant’s self reflection response. Finally, there is a user message containing the JSON of the UI being evaluated, instructions for evaluating it, and explicit instructions for the LLM to avoid common errors. There is an arrow showing that the LLM heuristic evaluation response goes into the LLM rephrasing prompt, which contains system instructions to rephrase the guideline violations into constructive feedback and the violations found from the LLM heuristic evaluation call.}
\end{figure*}

\end{document}